\documentclass[a4paper,twocolumn, iop, numberedappendix]{oja}
\usepackage{graphicx} 
\usepackage[T1]{fontenc}
\usepackage[utf8]{inputenc}
\usepackage{natbib}
\usepackage{fontawesome5}
\usepackage{multirow}
\usepackage{newtxtext,newtxmath}
\usepackage{comment}
\usepackage{graphicx}	
\usepackage{amssymb}	
\usepackage{xspace}        
\usepackage[dvipsnames]{xcolor}
\usepackage{booktabs}
\usepackage{tensor}
\usepackage{caption}
\usepackage{upgreek}
\usepackage{microtype}
\usepackage[colorlinks,linkcolor=blue,citecolor=blue,urlcolor=blue ]{hyperref}
\usepackage{xurl} 

\newcommand{\review}[1]{\textcolor{black}{#1}}
\newcommand{\rereview}[1]{\textcolor{black}{#1}}

\usepackage{cleveref}

\begin{document}
\title{A first measurement of baryonic feedback with Fast Radio Bursts}
\author{Robert Reischke$^{\dagger}$}
\affiliation{Argelander-Institut für Astronomie, Universität Bonn, Auf dem Hügel 71, D-53121 Bonn, Germany}
\thanks{$^\dagger$\href{mailto:reischke@posteo.net}{reischke@posteo.net}\\ \phantom{c$\!\mkern-1.5mu$} \href{mailto:rreischke@astro.uni-bonn.de}{rreischke@astro.uni-bonn.de}}

\author{Steffen Hagstotz}
\affiliation{Universitäts-Sternwarte, Fakultät für Physik, Ludwig-Maximilians Universität München, 
Scheinerstraße 1, D-81679 München, Germany and\\
Excellence Cluster ORIGINS, Boltzmannstraße 2, D-85748 Garching, Germany}

\begin{abstract}
\review{
Baryonic feedback fundamentally alters the total matter distribution on small to intermediate cosmological scales, posing a significant challenge for contemporary cosmological analyses. Direct tracers of the baryon distribution are therefore key for unearthing cosmological information buried under astrophysical effects. Fast Radio Bursts (FRBs) have emerged as a novel and direct probe of baryons, tracing the integrated ionised electron density along the line of sight, quantified by the dispersion measure (DM). The scatter of the DM as a function of redshift provides insight into the lumpiness of the electron distribution and, consequently, baryonic feedback processes.
}
\review{
Using a model calibrated to the \texttt{BAHAMAS} hydrodynamic simulation suite, we forward-model the statistical properties of the DM as a function of redshift. Applying this model to approximately 100 localised FRBs, we robustly constrain the governing feedback parameter, $\log  T_\mathrm{AGN} = {7.87^{+\hspace{.036cm}0.16}_{-0.22}}$ at 68\% confidence. Our findings represent the first measurement of baryonic feedback using FRBs, relating to power spectrum suppression and demonstrating a strong rejection of no-feedback scenarios at greater than 99.7\% confidence ($3\sigma$), depending on the FRB sample. We find that FRBs prefer fairly strong feedback, consistent with other measurements of the baryon distribution, as inferred from the thermal and kinetic Sunyaev-Zel'dovich effects. }

\review{
The results are robust against sightline correlations and modelling assumptions. We emphasise the importance of accurate calibration of the host galaxy and Milky Way contributions to the DM. Furthermore, we discuss implications for future FRB surveys and necessary improvements to current models to ensure accurate fitting of upcoming data, particularly that from low-redshift FRBs.
}
\end{abstract}

\maketitle
\section{Introduction}
\review{The ability to investigate small-scale cosmological structures is fundamentally limited by uncertainties in baryonic feedback.} Dynamic processes driven by active galactic nuclei (AGN), supernovae, and stellar winds redistribute gas both within and around haloes, thereby altering the overall matter distribution in the Universe and changing its statistical properties in the non-linear regime \cite{Rudd_pkSuppression_2008, VanDaalen2011, Chisari_Pk2019, VanDaalen_Pk2020, Schaye2023_flamingo}. These phenomena pose substantial modelling challenges for weak lensing measurements done by Euclid\footnote{\url{https://www.esa.int/Science_Exploration/Space_Science/Euclid}}, the Rubin Observatory Legacy Survey of Space and Time (LSST)\footnote{\url{https://www.lsst.org/}} and the Roman Space Telescope\footnote{\url{https://roman.gsfc.nasa.gov/}}, which depend on a precise understanding of matter clustering across a vast range of scales. 
A common approach to address these uncertainties is to implement scale cuts, thereby excluding small-scale data \citep[e.g.][]{2020A&A...634A.127A,Amon_DES_scalecut_2022, Secco_DES_scalecut_2022,des/kids:2023,wright_kids_2025}. However, such measures inevitably diminish the constraining power of the observations. To fully exploit the potential of small-scale cosmology, accurate models are needed to effectively marginalise over baryonic effects.
Hydrodynamical simulations offer a direct perspective on the interactions between baryons and dark matter \citep[see][for relevant works and overviews]{Dubois_HorizonAGN_2014,somerville_physical_2015,LeBrun_CosmoOwls_2014,Dolag_Magneticum_2016,McCarthy_BAHAMAS_2017, Springel_TNG_2018,vogelsberger_cosmological_2020,Schaye2023_flamingo}, but they require assumptions about the impact of sub-grid astrophysical processes that have to be calibrated to observations. 

Any external measurement of the baryon distribution is therefore key. Currently, the kinetic (kSZ) and thermal (tSZ) Sunyaev-Zel’dovich (SZ) effect \citep{sunyaev_observations_1972,sunyaev_microwave_1980} have been used in \citet{Schaan_velRec_2016,Hand:2012ui,Planck:2015ywj,DES:2016umt,Schaan2021_sz, Amodeo2021_sz,2022A&A...660A..27T,2025arXiv250707991K} to constrain feedback and can break degeneracies \citep[see also a discussion in][]{nicola_breaking_2022}. These measurements, in combination with X-ray data, are beginning to discriminate between different feedback models \citep{Schneider2022_WL_xray_kSZ, Bigwood2024, Grandis2023,2024arXiv240707152H,McCarthy_kSZ2024} and hint at stronger feedback.

A commonly used feedback prescription was put forward by \citet{2020A&A...641A.130M} (\textsc{HMCode2020}) by performing a modification to the concentration-mass relation of the
halo profiles, related to the strength of the AGN feedback
This feedback model is then calibrated
against a suite of hydrodynamical simulations from \texttt{BAHAMAS} \citep{McCarthy_BAHAMAS_2017,2018MNRAS.476.2999M}, generating a one-parameter family of feedback models covering a wide range of cosmologies. This model has been used in several weak lensing analyses \citep[see e.g.][]{asgari_kids_2021,amon/etal:2022,2022A&A...660A..27T,wright_kids_2025,2025arXiv250319442S} to marginalise over or to constrain feedback. In their combined analysis, \citet{des/kids:2023} showed that the accuracy of this model is good enough to yield unbiased cosmological constraints compared to other frequently used emulators \citep[e.g.][]{arico/etal:2021b,knabenhans/etal:2023}. 

A novel probe to constrain feedback is Fast Radio Bursts (FRBs), which were first discovered in archival data \citep{2007Sci...318..777L}. Their origin is still debated \citep{2019A&ARv..27....4P}, and possibilities range from Magnetars \citep{2013Sci...341...53T,2020Natur.587...59B} to binary mergers \citep{2016ApJ...826...82L}.
FRBs are short, transient events lasting only a few milliseconds, with frequencies ranging from approximately 100 MHz to several GHz.
The initial pulse is dispersed due to free electrons in the ionised galactic and intergalactic medium (IGM), leading to a delayed arrival time of the pulse frequencies, $\Delta t(\nu) \propto \nu^{-2}$. The proportionality constant is appropriately called the dispersion measure (DM) and is directly related to the integrated electron density along the line of sight.
The key observable of FRBs of cosmological interest is therefore the DM-$z$ relation, also known as the Macquart relation. This has been applied to current data to measure the baryon density and the Hubble constant \citep[e.g.][]{2014PhRvD..89j7303Z,2020Natur.581..391M,2022MNRAS.511..662H,2022MNRAS.516.4862J,2022MNRAS.515L...1W, 2024arXiv241007084K,2025JCAP...01..036A,2025arXiv250706841Z}.
In addition, the statistical properties of the DM are an excellent probe of the statistical properties of the LSS if enough FRBs are available \citep[e.g.][]{mcquinn_locating_2014,masui_dispersion_2015,rafiei-ravandi_chimefrb_2021,bhattacharya_fast_2021,2021MNRAS.502.2615T,2021PhRvD.103b3517R,2022MNRAS.512..285R,2023arXiv230909766R,2025OJAp....8E.127R, 2025arXiv250418745S, 2025arXiv250604186H,2025ApJ...983...46M,2025A&A...698A.163H}. 

In this paper, we use localised FRBs, i.e., events for which a host galaxy and therefore an independent redshift estimate have been found, to constrain the scatter of the Macquart relation. Baryonic feedback smoothes out the distribution of baryons relative to the dark matter distribution. Therefore, strong feedback will lead to less scatter in the DM-$z$ relation, whereas weak feedback will lead to greater scatter. We leverage this signal \citep[see][]{reischke_cosmological_2023,2025arXiv250418745S} to put constraints on $\log T_\mathrm{AGN}$ directly linked to the power spectrum suppression. We can recover this small signal by imposing a tight prior on the cosmological parameters, as measured by \citet{planck_collaboration_planck_2020}. 

We organise the paper as follows: In Section \ref{sec:methodology} we introduce the methodology, including the DM and its statistical properties and how they relate to feedback, in particular in the \texttt{BAHAMAS} model. Additionally, we describe the statistical model used and how the different components of the DM enter the final likelihood. Section \ref{sec:results} presents the data, the selections applied, and the resulting constraints on baryonic feedback. Furthermore, we investigate the model's goodness of fit when analysed using different subsets of the data. Lastly, we present the suppression of the matter power spectrum. In Section \ref{sec:conclusions}, we conclude and point out future directions. The FRB sample used is available in Appendix \ref{app:data}.

\section{Methodology}
\label{sec:methodology}
In this section, we describe how the DM relates to the cosmological electron density and derive a corresponding likelihood for the Macquart relation. The material here closely follows that presented in \citet{reischke_cosmological_2023}. In particular, we review the dispersion measure in Section \ref{subsec:dm}, its statistical properties in Section \ref{subsec:cov} and the different ingredients of the likelihood in Section
\ref{subsec:likelihood}.
\subsection{Dispersion measure}
\label{subsec:dm}
The total DM, $\mathcal{D}^{\mathstrut}_\mathrm{tot}$, of an FRB at the sky position $\hat{\boldsymbol{x}}$ and redshift $z$ is given by\footnote{In the literature, it is common to write the dispersion measure as DM also in equations. We find that this impairs the readability of expressions; we therefore use $\mathcal{D}$.}
\begin{equation}
\label{eq:dmtot}
    \mathcal{D}^{\mathstrut}_\mathrm{tot}(\hat{\boldsymbol
    {x}}, z) = \mathcal{D}^{\mathstrut}_\mathrm{host}(z) + \mathcal{D}^{\mathstrut}_\mathrm{MW}(\hat{\boldsymbol
    {x}}) + \mathcal{D}^{\mathstrut}_\mathrm{LSS}(z,\hat{\boldsymbol
    {x}}) \; .
\end{equation}
The dispersion is associated with electrons associated with the host halo, with the Milky Way\footnote{As well as its halo, $\mathcal{D}^{\mathstrut}_\mathrm{MW}\to\mathcal{D}^{\mathstrut}_\mathrm{MW} + \mathcal{D}^{\mathstrut}_\mathrm{halo}$.}, or with the large-scale structure (LSS). 
Note that all of these contributions are random variables with their own associated probability density functions. Fitting the Macquart relation is concerned with finding the expectation value of \Cref{eq:dmtot}, or more precisely, the spatial average (equivalent to the ensemble average).
Let us first focus on the LSS contribution
\begin{equation}
\label{eq:DM_LSS}
    \mathcal{D}^{\mathstrut}_\mathrm{LSS}(\hat{\boldsymbol
    {x}},z) = c\int_0^z \! n^{\mathstrut}_\mathrm{e}(\hat{\boldsymbol
    {x}},z') \, f^{\mathstrut}_\mathrm{IGM}(z') \, \frac{1+z'}{H(z')} \, \mathrm d z' \; ,
\end{equation}
where $n_\mathrm{e}(\hat{\boldsymbol
    {x}},z)$ is the {comoving} cosmic free electron density, $c$ is the speed of light, $H(z) = H_0 E(z)$ is the Hubble function with the expansion function $E(z)$ and the Hubble constant $H_0$ and $f^{\mathstrut}_\mathrm{IGM}(z)$ is the fraction of free electrons in the intergalactic medium\footnote{We use \href{https://github.com/FRBs/FRB}{https://github.com/FRBs/FRB} to calculate this contribution and refer to \citet{2020Natur.581..391M} for details.}. 
\review{We calculate $f_\mathrm{IGM}$ by subtracting the fraction bound in stars, compact objects and the cold interstellar medium (ISM):
\begin{equation}
\label{eq:f_IGM}
    f^{\mathstrut}_\mathrm{IGM}(z) = 1 - f^{\mathstrut}_\star(z) - f^{\mathstrut}_\mathrm{ISM}(z) \, .
\end{equation}
\rereview{The original model found $f_\mathrm{IGM}(z)|_{z=0} = 0.84$ which slowly increases towards higher redshifts. More recent values show, however, that $f_\mathrm{IGM}(z)|_{z=0} \approx 0.92$ \citep[see e.g.][]{2019MNRAS.489.1619H,2025arXiv251119620C}. We will keep the redshift dependence fixed, marginalise over the amplitude at redshift zero and show that the influence on the feedback constraints is negligible (see Section \ref{subsec:limitations}).
In the future, one could also try to model the fraction of hot (ionised) gas and leave it free \citep[see e.g.][]{2025arXiv250707892S}}.} Next, we express the electron density via the number of baryons in the Universe:
\begin{align}
\label{eq:n_e}
    n^{\mathstrut}_\mathrm{e}(\hat{\boldsymbol
    {x}},z) =  \chi^{\mathstrut}_\mathrm{e} \frac{\rho^{\mathstrut}_\mathrm{b}(\hat{\boldsymbol
    {x}}, z)}{m_\mathrm{p}} =\chi^{\mathstrut}_\mathrm{e} \frac{\bar \rho_\mathrm{b}}{m_\mathrm{p}} \big[ 1 + \delta^{\mathstrut}_\mathrm{e} (\hat{\boldsymbol
    {x}}, z) \big]\;,
\end{align}
with the proton mass $m_\mathrm{p}$ and the baryon density $\rho^{\mathstrut}_\mathrm{b}$. The electron fraction is given by
\begin{align}
\label{eq:chi_e}
    \chi^{\mathstrut}_\mathrm{e} &= Y^{\mathstrut}_\mathrm{H} + \frac{1}{2} Y^{\mathstrut}_\mathrm{He} \\
    & \approx 1 - \frac{1}{2} {Y}^{\mathstrut}_\mathrm{He} \, ,
\end{align}
and can be calculated from the primordial hydrogen and helium abundances $Y_\mathrm{H}$ and $Y_\mathrm{He}$. Here, we assume $Y_\mathrm{H} \approx 1 - Y_\mathrm{He}$ and $Y_\mathrm{He} = 0.245$, found to high precision by CMB measurements \citep{planck_collaboration_planck_2020} and spectroscopic observations of metal-poor gas clouds \citep{2015JCAP...07..011A}. The baryon density in \Cref{eq:n_e} can be expanded into a background contribution, $\bar \rho_\mathrm{b}$ and perturbations to it generated by the electron density contrast, $\delta_\mathrm{e}$, such that the expectation value is $\langle\delta_\mathrm{e}\rangle = 0$. 

Putting all components together, \Cref{eq:DM_LSS} turns into
\begin{equation}
\label{eq:DM_LSS_v2}
    \mathcal{D}^{\mathstrut}_\mathrm{LSS}(\hat{\boldsymbol{x}},z) = \mathcal{A}  \int_0^z   \frac{1+z'}{E(z')} f^{\mathstrut}_\mathrm{IGM} (z^\prime) \big[1+\delta^{\mathstrut}_\mathrm{e}(\hat{\boldsymbol{x}},z')\big] \mathrm{d} z' ,
\end{equation}
where we defined the overall amplitude
\begin{equation}
    \mathcal{A}:= \frac{3 \Omega_\mathrm{b0} \chi_\mathrm{H}}{8 \uppi G m_\mathrm{p}}\chi_\mathrm{e}
\end{equation}
with the dimensionless baryon density parameter $\Omega_\mathrm{b0}$ today and the Hubble radius, $\chi_\mathrm{H} = c/H_0$.

Averaging \Cref{eq:DM_LSS_v2} recovers the known DM-redshift relation \citep[see e.g.][]{ioka_cosmic_2003,inoue_probing_2004, deng_cosmological_2014}
\begin{equation}
\label{eq:DM_LSS_avg}
    \langle\mathcal{D}^{\mathstrut}_\mathrm{LSS}(\hat{\boldsymbol{x}},z)\rangle = \mathcal{A} \,  \int_0^z \,f^{\mathstrut}_\mathrm{IGM}(z^\prime) \frac{1+z'}{E(z')} \mathrm{d} z' \; .
\end{equation}

\subsection{Covariance}
\label{subsec:cov}
We assume to have a set of $N_\mathrm{FRB}$ host-identified FRBs $\big\{\mathcal{D}_{\mathrm{obs},i},\hat{\boldsymbol{x}}_i, z^{\mathstrut}_i\big\}$, $i= 1,...,N_\mathrm{FRB}$, with the observed DM, the direction of the burst and its redshift. The covariance induced by the LSS (i.e. induced by cosmic variance) of these bursts is given by \citep{reischke_cosmological_2023}\footnote{The code for the covariance is publicly available via \href{https://github.com/rreischke/frb_covariance}{https://github.com/rreischke/frb\_covariance}.}:
\begin{equation}
\label{eq:final_covariance}
        {\mathrm{C}}^{\mathstrut}_{ij} = 
      \sum_\ell \frac{2\ell+1}{4\pi}P^{\mathstrut}_\ell\left(\cos \theta^{\mathstrut}_{ij}\right)C^{\mathcal{DD}}_{ij}(\ell,z)\;, 
\end{equation}
with the Legendre polynomials $P_\ell(x)$ and the angular separation between pairs of FRBs given via $\hat{\boldsymbol{x}}_i\cdot \hat{\boldsymbol{x}}_j = \cos \theta_{ij}$. Lastly, $C_{ij}^{\mathcal{DD}}(\ell,z)$ is the angular power spectrum of the dispersion measure
\begin{equation}
   C_{ij}^{\mathcal{DD}}(\ell,z) =  \frac{2}{\pi}\int k^2 I(\ell,k, z^{\mathstrut}_i) I(\ell,k,z_j)\;\mathrm{d}k  \\ 
\end{equation}
with 
\begin{equation}
I(\ell,k,z) = \mathcal{A}\int_0^{z} f^{\mathstrut}_\mathrm{IGM}(z^\prime) \frac{1+z}{E(z)} \sqrt{P^{\mathstrut}_\mathrm{e}(k,z')} j_\ell [k \chi(z^\prime)]\, \mathrm{d}z'\;.
\end{equation}
Here, $j_\ell(x)$ are spherical Bessel functions, $\chi$ is the comoving distance, and $P^{\mathstrut}_\mathrm{e}(k,z)$ the electron power spectrum defined via
\begin{equation}
    \big\langle \delta_\mathrm{e}(\boldsymbol{k})\delta_\mathrm{e}(\boldsymbol{k}^\prime) \big\rangle = (2\pi)^3 \delta^{(3)}_\mathrm{D}(\boldsymbol{k} + \boldsymbol{k}^\prime)P^{\mathstrut}_\mathrm{e}(k,z)\;,
\end{equation}
with the Dirac delta distribution $\delta^{(3)}_\mathrm{D}$. We compute the electron power spectrum using the emulator trained by \citet{2025OJAp....8E..72N} with \texttt{cosmopower} \citep{2022MNRAS.511.1771S}. The emulator itself is based on a fit motivated by the halo model \citep{Mead2015,2020A&A...641A.130M,2022A&A...660A..27T} to the \texttt{BAHAMAS} simulations. It covers most cosmological parameter space important for LSS cosmology as well as the feedback strength varied in \texttt{BAHAMAS}, denoted by the parameter $\log T_\mathrm{AGN}$. The initial linear matter power spectrum has been calculated with \texttt{HiCLASS} \citep{2011JCAP...07..034B,2017JCAP...08..019Z}.

\subsection{Likelihood}
\label{subsec:likelihood}
The likelihood of the total DM, i.e. $\mathcal{D}_\mathrm{tot}$, is given by marginalising over the host contribution and the Milky Way contribution:
\begin{equation}
\label{eq:full_lh}
\begin{split}
     p\left(\mathcal{D}^{\mathstrut}_\mathrm{tot}\big|z\right) =&\; \int\mathrm{d}\mathcal{D}^{\mathstrut}_\mathrm{host}\int \mathrm{d}\mathcal{D}^{\mathstrut}_\mathrm{MW}
     \int\mathrm{d}\mathcal{D}^{\mathstrut}_\mathrm{halo}\\
     &\times p^{\mathstrut}_\mathrm{MW}(\mathcal{D}^{\mathstrut}_\mathrm{MW}) p^{\mathstrut}_\mathrm{halo}(\mathcal{D}^{\mathstrut}_\mathrm{halo}) p^{\mathstrut}_\mathrm{host}(\mathcal{D}^{\mathstrut}_\mathrm{host})\\&\times p^{\mathstrut}_\mathrm{LSS}\left(\mathcal{D}^{\mathstrut}_\mathrm{tot} - \frac{\mathcal{D}^{\mathstrut}_\mathrm{host}}{1+z} - \mathcal{D}^{\mathstrut}_\mathrm{MW} - \mathcal{D}^{\mathstrut}_\mathrm{halo}\bigg|z^{\mathstrut}_i\right)\;.
\end{split}
\end{equation}
The host DM, $\mathcal{D}_\mathrm{host}$, is the DM in the rest-frame of the host and therefore gets redshifted by a factor $(1+z)$ when included in $p_\mathrm{LSS}$.
The total likelihood is then given, assuming that all FRBs are independent (we will discuss this in more detail in Section \ref{subsec:correlations}), by:
\begin{equation}
\label{eq:likelihood:full}
    L = \prod_{i=1}^{N_\mathrm{FRB}}p\left(\mathcal{D}^{\mathstrut}_\mathrm{obs,i}\big|z^{\mathstrut}_i\right)\;.
\end{equation}
In the remaining part of this section, we will discuss the specific ingredients entering \Cref{eq:full_lh}.

\subsubsection{LSS dispersion measure PDF}
\review{There is strong evidence that the DM one-point probability density function (PDF) can be reasonably well modelled by a log-normal distribution \citep[e.g.][]{mcquinn_locating_2014,2025arXiv250418745S}. However, to date, it has been common practice to use the fitting formula provided by \citet{2000ApJ...530....1M}. \citet{2025arXiv250418745S} discussed why this distribution has its shortcomings and is not an actual good fit to hydrodynamical simulations \citep{zhang_intergalactic_2021,2024A&A...683A..71W} and that the PDF is indeed better described by a log-normal distribution
\begin{equation}
    \label{eq:ln_general}p^{\mathstrut}_\mathrm{ln}\left(x;\mu^{\mathstrut}_\mathrm{ln},\sigma^{\mathstrut}_\mathrm{ln}\right) = \frac{1}{x\sigma^{\mathstrut}_\mathrm{ln}\sqrt{2\pi}}\exp\left[-\frac{\left(\log x - \mu^{\mathstrut}_\mathrm{ln}\right)^2}{2\sigma^2_\mathrm{ln}}\right]\;.
\end{equation}
That said, \citet{2025arXiv250707090K} recently observed deviations from a log-normal distribution, particularly at redshifts. For the sample size analysed here, differences in the functional form of the likelihood will not translate into biased parameter constraints. This is, for example, shown in \citet{2026arXiv260118784T}, who derived an analytical form of the probability distribution function from gas profiles and a more complex feedback model.
}
The average of the log-normal, $\mu_\mathrm{ln}$, for an FRB $i$, can be related to the cosmological mean, $\mathcal{D}_{\mathrm{LSS,}i} \equiv \langle\mathcal{D}_\mathrm{LSS}(z_i)\rangle$ in \Cref{eq:DM_LSS_avg}, and its variance, $\mathrm{C}_{ii}$ calculated from the diagonal elements of \Cref{eq:final_covariance}:
\begin{equation}
\label{eq:ln_mu}
    \mu_{\mathrm{ln},i} = \log\left(\frac{\mathcal{D}^{\mathstrut 2}_{\mathrm{LSS,}i}}{\sqrt{\mathcal{D}^{\mathstrut 2}_{\mathrm{LSS,}i}  + {\mathrm{C}^{\mathstrut}_{ii}}}}\right)\;.
\end{equation}
Likewise, the log-normal variance is
\begin{equation}
\label{eq:ln_sigma}
    \sigma^2_{\mathrm{ln},i} = \log\left(1+\frac{{\mathrm{C}}_{ii}}{\mathcal{D}^{\mathstrut 2}_{\mathrm{LSS,}i}}\right)\;,
\end{equation}
so that:
\begin{equation}
    p^{\mathstrut}_\mathrm{LSS}\left(\mathcal{D}^{\mathstrut}_\mathrm{LSS}\big|z^{\mathstrut}_i\right) = p^{\mathstrut}_\mathrm{ln}\left(\mathcal{D}^{\mathstrut}_\mathrm{LSS};\mu^{\mathstrut}_{\mathrm{ln,}i},\sigma^{\mathstrut}_{\mathrm{ln},i}\right)\;.
\end{equation}

Note that the off-diagonal elements of \Cref{eq:final_covariance} describing correlations between different sightlines cannot be included in the case of a log-normal PDF. We address this limitation explicitly in Section \ref{subsec:correlations}.

The results for assuming a Gaussian distribution for $p_\mathrm{LSS}$ are shown in Appendix \ref{app:likelihood}. The inability to fit the long tail leads to an overestimation of the Gaussian variance and, consequently, an underestimation of the feedback strength.

\begin{figure}
    \centering
    \includegraphics[width=0.45\textwidth]{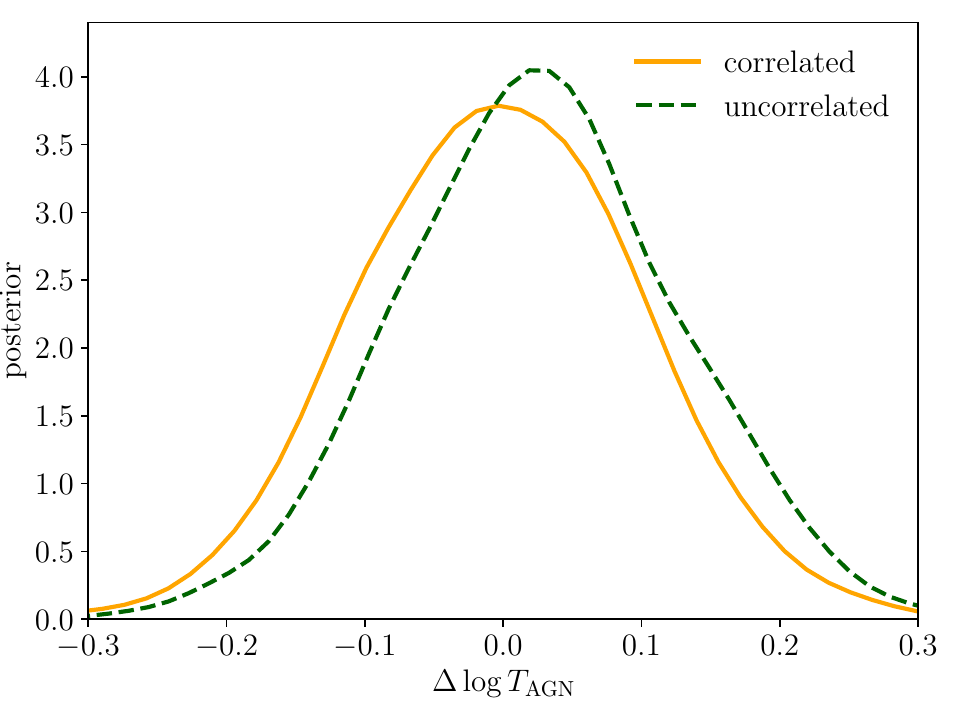}
    \caption{Influence of the correlation of the DM from different sightlines on the constraints on the feedback parameter $\log T_\mathrm{AGN}$ discussed in Section \ref{subsec:correlations}. The solid orange line shows the constraints in the Gaussian toy model if the covariance of all events is properly taken into account, while the green dashed line ignores any correlations.}
    \label{fig:correlation_test}
\end{figure}

\subsubsection{Host dispersion measure PDF}
The host contribution as measured from hydrodynamic simulations follows a log-normal distribution \citep{2020AcA....70...87J,2024ApJ...967...32M,2024arXiv240308611T}; see also \citet{mcquinn_locating_2014,2025OJAp....8E.127R} for more theoretical approaches. We will therefore assume the same functional form as in \Cref{eq:ln_general} for $p_\mathrm{host}$ and use \Cref{eq:ln_mu,eq:ln_sigma} to calculate the corresponding log-normal mean and variance derived from the mean host contribution, $\langle\mathcal{D}_\mathrm{host}\rangle$, and its variance, $\sigma^2_\mathrm{host}$ which we will treat as free parameters for the host contribution in our analysis:
\begin{equation}
p^{\mathstrut}_\mathrm{host}\left(\mathcal{D}^{\mathstrut}_\mathrm{host}\right) = p^{\mathstrut}_\mathrm{host}\left(\mathcal{D}^{\mathstrut}_\mathrm{host};\mu^{\mathstrut}_{\mathrm{ln}},\sigma^{\mathstrut}_{\mathrm{ln}}\right) \; .
\end{equation}
We do not assume the rest-frame host properties to vary significantly over the redshift range used in this analysis, consistent with results from numerical simulations \citep{2024arXiv240308611T}.

\subsubsection{Milky Way dispersion measure PDF}
The Milky Way contribution can be split into two parts: the interstellar medium (ISM) contribution and the extended halo contribution. The distinction is useful, as the former can be modelled fairly well at high galactic latitudes \citep{2020ApJ...897..124O} using galactic Pulsar dispersion measures. We use the \texttt{python} implementation \citep{2024RNAAS...8...17O} of the model introduced in \citet{2002astro.ph..7156C} to obtain the mean contribution from the ISM and assume a Gaussian $p_\mathrm{MW}$ model with a 10\% relative error. \citet{2017ApJ...835...29Y} introduced an alternative model for the ISM contribution; the results here, however, do not depend on this choice \citep[see e.g.][]{2023MNRAS.523.6264R}.
The halo part, however, is less constrained, and can have varying contributions \citep[as demonstrated for example in][]{2019MNRAS.485..648P,2020ApJ...888..105Y,2023ApJ...946...58C}, showing values of up to $100\;\mathrm{pc}\;\mathrm{cm}^{-3}$. These values, however, might still have contributions from the galactic ISM component. We therefore assume a mean halo contribution
$\langle\mathcal{D}_\mathrm{halo}\rangle =50\;\mathrm{pc}\;\mathrm{cm}^{-3}$ and marginalise over $\mathcal{D}_\mathrm{halo}$ with a Gaussian $p_\mathrm{halo}$ distribution and a 20\% relative error budget.

\subsection{Posterior}
The parameter set analysed here is given by the following vector
\begin{equation}
    \boldsymbol{\theta} = \big(\langle\mathcal{D}_\mathrm{host}\rangle,\sigma_\mathrm{host},\log T_\mathrm{AGN},\!\underbrace{\Omega_\mathrm{m},\Omega_\mathrm{b},\sigma_8,n_\mathrm{s},h}_{\mathrm{cosmological\;parameters}} \!\!\big)^{\top}\;,
\end{equation}
which enter the likelihood via $p_\mathrm{LSS}$ and $p_\mathrm{host}$. Here, $\sigma_8$ controls the amplitude of the linear matter power spectrum and hence the amplitude of the electron power spectrum. $n_\mathrm{s}$ is the spectral index, determining the slope of the initial power spectrum. For the cosmological parameters, we assume a correlated prior based on the results from \citet{planck_collaboration_planck_2020} by resampling the public $\Lambda$CDM Monte Carlo Markov Chains (MCMCs). Effectively, this fixes the cosmological parameters at their best-fit values, and the remaining uncertainty has little impact on our analysis. Given the likelihood, \Cref{eq:likelihood:full}, the posterior is
\begin{equation}
    p\left(\boldsymbol{\theta}\big|\big\{\mathcal{D}_{\mathrm{obs},i},\hat{\boldsymbol{x}}_i, z^{\mathstrut}_i\big\}\right) \propto L\left(\big\{\mathcal{D}_{\mathrm{obs},i},\hat{\boldsymbol{x}}_i, z^{\mathstrut}_i\big\}\big|\boldsymbol{\theta}\right)p^{\mathstrut}_\mathrm{planck}(\boldsymbol{\theta})\;.
\end{equation}
We use \texttt{emcee} \citep{2013PASP..125..306F} to sample the posterior; the parameters and their priors are summarised in \Cref{tab:parameters_priors}.

\renewcommand{\arraystretch}{1.5}
\begin{table}[]
    \centering
    \begin{tabular}{ccr}
    \hline\hline
        parameter &  prior & interpretation  \\ \hline
        $\langle \mathcal{D}_\mathrm{host}\rangle$ & $[10,\infty)$ &mean host contribution\\
        $\sigma_\mathrm{host}$ & $[10,\infty)$ &scatter in the host contribution \\
         $\log T_\mathrm{AGN}$ & $[7,8.6]$ & feedback strength \\ \hline
         $\Omega_\mathrm{m0}$ & Planck& matter density parameter\\ 
         $\Omega_\mathrm{b0}$ &Planck& baryon density parameter\\
         $\sigma^2_8$ &Planck& linear power spectrum variance at $8\;\mathrm{Mpc}/h$\\
         $n_\mathrm{s}$ &Planck& linear power spectrum slope at large scales\\
         $h$ &Planck & dimensionless Hubble constant
    \end{tabular}
    \caption{Parameters used in the analysis here with their respective prior values and a brief explanation. Planck refers to the correlated priors from the public chains \citep{planck_collaboration_planck_2020}.}
    \label{tab:parameters_priors}
\end{table}

\section{Results}
\label{sec:results}
This section shows the main results of this paper. We briefly discuss the data and selection cuts in Section \ref{sec:data}. In Section \ref{subsec:correlations}, we analyse correlations across different sightlines and further reduce the data set. The constraints on feedback in the form of $\log T_\mathrm{AGN}$ are shown in Section \ref{subsec:constraints}, and we discuss the corresponding goodness-of-fit in Section \ref{subsec:gof}. \Cref{fig:suppression} in Section \ref{subsec:sup} shows the matter power spectrum suppression as measured by FRBs, the main result of this work.

\begin{figure}
    \centering
    \includegraphics[width=0.47\textwidth]{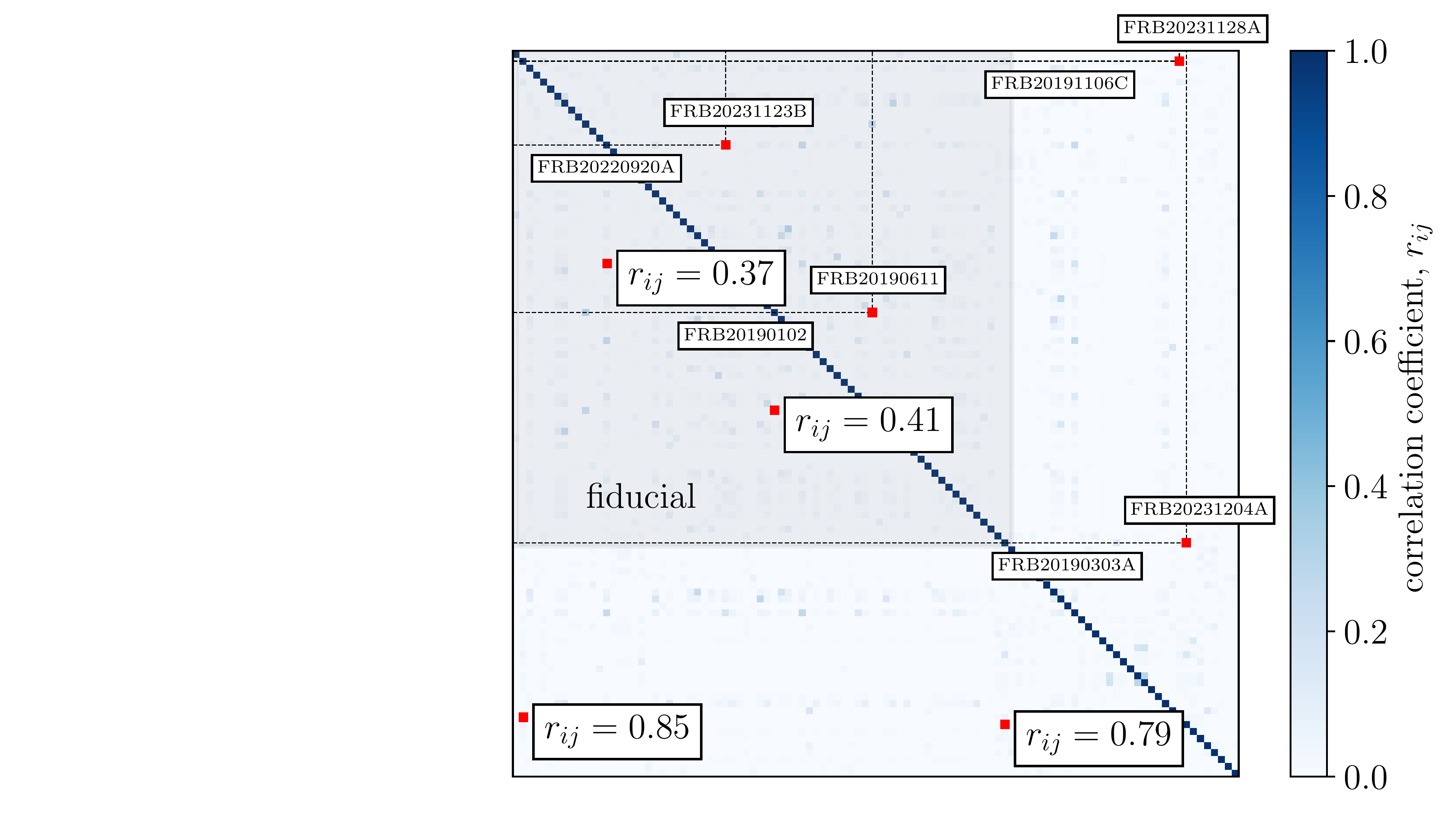}
    \caption{Shown is the correlation coefficient, $r_{ij}$, with pairs, $(i,j)$, with $r_{ij} > 0.3$ highlighted in red. The horizontal dashed lines mark the corresponding pair of FRBs. The small white boxes in the upper half mark the name of the corresponding correlated FRBs, while the white boxes in the lower half indicate the actual correlation coefficient of each labelled FRB pair. The grey shaded area shows the fiducial data set from \Cref{tab:app_frb_list}.}
    \label{fig:correlation_matrix}
\end{figure}

\subsection{Data}
\label{sec:data}
We list the FRBs with identified host redshifts used in this analysis in \Cref{tab:app_frb_list,tab:app_frb_list_2} in Appendix \ref{app:data}. 
Our fiducial data set are the events shown in \Cref{tab:app_frb_list} from which we remove two FRBs, FRB$\,$20190520B and FRB$\,$20220831A due to their high DM at very low redshift, indicating strong and, cosmologically speaking, local dispersion. Removing these sources does not affect our results, provided that the local dispersion is uncorrelated with the LSS. The same procedure was applied in \citet{2025NatAs.tmp..131C} and the fiducial set used here is equivalent to theirs. The second set consists of additional FRBs which are primarily at very low redshift and are therefore strongly affected by the Milky Way and the FRB host galaxy. We refer to the combination of these as the full set.

\begin{figure}
    \centering
    \includegraphics[width=0.48\textwidth]{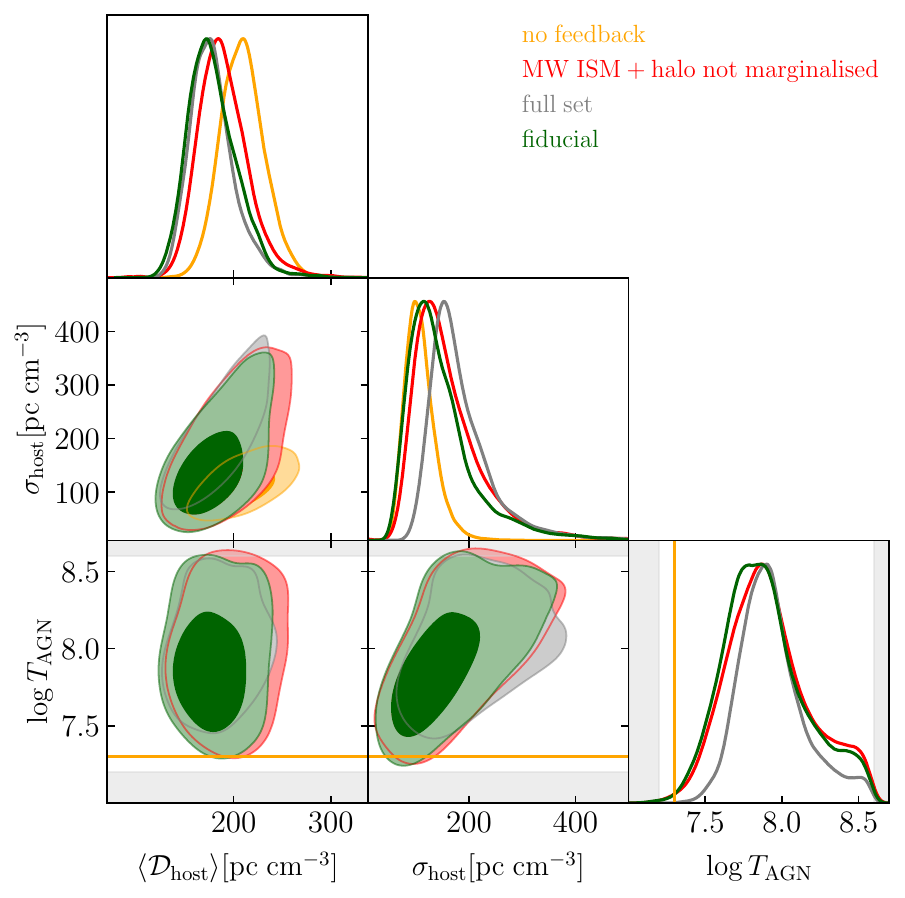}
    \caption{Confidence intervals (68\% and 95\%) on the three parameters of interest, the host DM, $\langle\mathcal{D}_\mathrm{host}\rangle$, its variance, $\sigma_\mathrm{host}$ and the feedback strength $\log T_\mathrm{AGN}$. The orange vertical and horizontal lines indicate the value of $\log T_\mathrm{AGN}$ at which there is no feedback, i.e. the total matter distribution follows the dark matter distribution. In green, we show the contours of our fiducial run, while the orange contours assume no feedback ($\log T_\mathrm{AGN}= 7.3$). The red contours show the impact of not marginalising over the MW and the host contribution. The grey contour shows the full set of FRBs (with the selection criterion discussed in Section \ref{subsec:correlations}). Lastly, the light grey-shaded region indicates the prior range.}
    \label{fig:corner}
\end{figure}

\subsection{Correlations}
\label{subsec:correlations}
First, we assess the impact of correlations between different sightlines arising from LSS. To this end, we assume, for this test, a Gaussian likelihood with covariance matrix $\boldsymbol{\mathrm{C}}_\mathrm{tot}$ such that:
\begin{equation}
    \boldsymbol{\mathrm{C}}_\mathrm{tot} = \boldsymbol{\mathrm{C}}_\mathrm{LSS} + \boldsymbol{\mathrm{C}}_\mathrm{MW} + \boldsymbol{\mathrm{C}}_\mathrm{host}\;,
\end{equation}
where the elements of $\boldsymbol{\mathrm{C}}_\mathrm{LSS}$ are given by \Cref{eq:final_covariance}. For the MW and the host, we assume a diagonal covariance with standard deviations of $30\mathrm{\,pc\;cm}^{-3}$ and $\sigma_\mathrm{host}$ as a free parameter rescaled by the host redshift. Although this assumes the wrong shape of the likelihood, it will give us an estimate of how large the effect of neglecting the correlations in the data set is. Recall that we assume all sightlines to be independent in \Cref{eq:likelihood:full}, since it is not possible to incorporate the correlations self-consistently for a log-normal distribution without running forward simulations \citep{2024arXiv241007084K}. The result of this exercise is displayed in 
\Cref{fig:correlation_test}, where we can see that we get slightly larger values of $\log T_\mathrm{AGN}$ if we ignore the covariance between different events. Note that we only show relative values here, as the exact value of the result will not be correct in the first place due to the Gaussian assumption of the likelihood.
While the effect is not statistically significant, we investigate this result further. In \Cref{fig:correlation_matrix}, the correlation coefficient, $r_{ij}$, of the covariance matrix with those pairs of FRBs where it is larger than 0.3 is marked in red\footnote{A selection based on correlation can also remove repeating bursts. While this sounds trivial, it is notoriously difficult to clean one's catalogue from repeating bursts due to the lack of a unified catalogue between or even within the same instrument.}. We exclude 4 FRBs to reduce the correlation in the data set. Here, we always remove the lower-redshift object to preserve more cosmological information. The selection is based on redshift so that we do not preferentially select objects with a larger DM.

Using $r_{ij} > 0.3$ as an arbitrary threshold will leave some degree of correlation in the data set. In fact, using $r_{ij} > 0.2$ already results in 16 correlated pairs. Given the currently available sample size, the effect remains small, and we continue the analysis with the correlation cut outlined above. Nonetheless, this again highlights the importance of correlations between various FRBs moving forward \citep[see][for more details]{reischke_cosmological_2023}.

\subsection{Constraints on feedback}
\label{subsec:constraints}
We first present our fiducial constraints, i.e. using the selection discussed in Section \ref{sec:data} and \ref{subsec:correlations} and only the FRBs shown in \Cref{tab:app_frb_list}. The resulting posterior contours are shown in \Cref{fig:corner} in green. As a comparison, the results without any feedback are shown in orange, which corresponds to $\log T_\mathrm{AGN}= 7.3$ \citep{des/kids:2023}. Anything with lower values of $\log T_\mathrm{AGN}$ would correspond to an increase in the power spectrum relative to a cold dark matter-only simulation. 

It can be seen that the fiducial results have almost zero probability mass for vanishing feedback, thereby rejecting the possibility that the free electrons follow the dark matter distribution. This agrees with the results of \citet{2025NatAs.tmp..131C}, indicating that feedback drives baryons out of galaxies and halos. Inspecting the green contours further, one can see that the host contribution compensates for the smaller electron clustering relative to the dark matter. Due to a larger $\log T_\mathrm{AGN}$, baryons get expelled more, and the electron distribution becomes smoother, decreasing its variance. In turn, the orange contour shows smaller host variance as a larger fraction can be accommodated by the LSS, which is clumpier due to the lack of feedback. Lastly, we indicate the prior boundaries by a light grey-shaded area.

\begin{figure}
    \centering
    \includegraphics[width=0.45\textwidth]{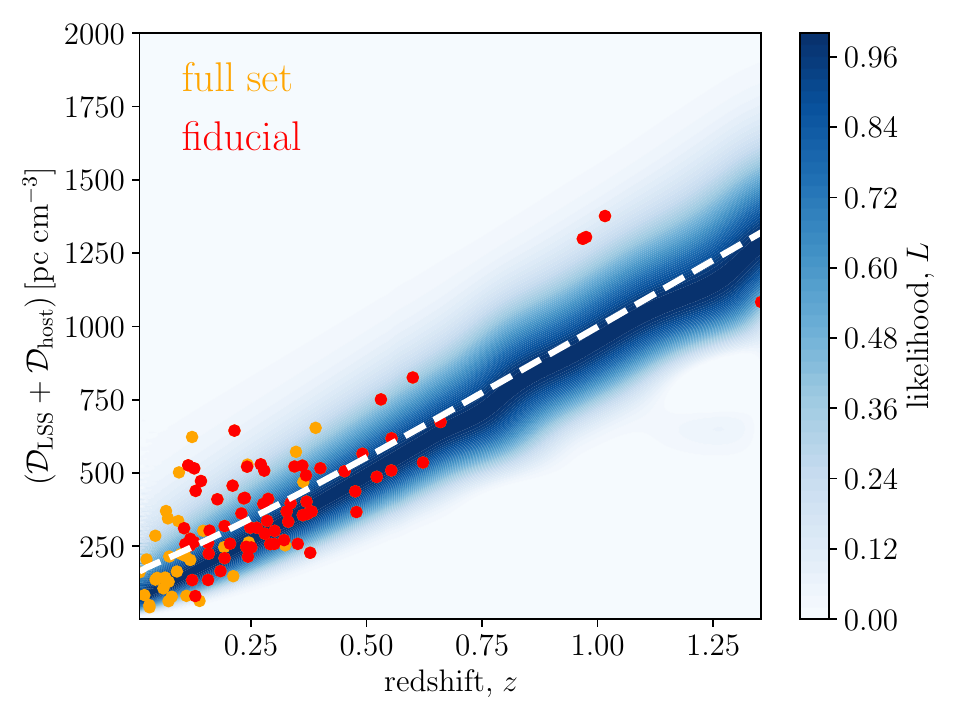}
    \caption{Excess DM, $\mathcal{D}^{\mathstrut}_\mathrm{ex}=\mathcal{D}^{\mathstrut}_\mathrm{LSS} + \mathcal{D}^{\mathstrut}_\mathrm{host}$, as a function of redshift, $z$. The red dots indicate the fiducial data set used. The mean of the excess DM is shown as the white dashed line. As an overlay, we show the likelihood normalised to unity at the peak at each redshift and evaluated at the best fit derived from the full data set. The orange dots indicate the additional FRBs that comprise the full set.}
    \label{fig:likelihood}
\end{figure}

We also check the impact of not marginalising over the Milky Way ISM and halo contribution and show the results in red (this corresponds to replacing $p_\mathrm{MW}$ and $p_\mathrm{halo}$ by delta distributions in Equation \ref{eq:full_lh}). It can be seen that the effect on the constraints is minimal (compare red with green) for the current data set. Ignoring the uncertainties associated with the Milky Way significantly accelerates the likelihood calculation.

Furthermore, we include the full set of FRBs into the analysis, i.e. \Cref{tab:app_frb_list,tab:app_frb_list_2} (again with the mentioned selection criteria in mind). Since the additional FRBs are located at low redshifts, they mostly constrain the host contribution, as it is the dominant component at redshifts $z\lesssim 0.5$ \citep[see the discussion in Section 6.1 of][]{2024arXiv241007084K}. The grey contour is significantly tighter compared to the green (fiducial) one for the host contribution. This, in turn, also frees up constraining power for the feedback parameter $\log T_\mathrm{AGN}$, comfortably ruling out a non-feedback scenario by more than 3$\sigma$.

In \Cref{fig:likelihood}, we show the best-fit likelihood for the excess DM $\mathcal{D}_\mathrm{LSS} + \mathcal{D}_\mathrm{host}$ as a function of redshift $z$, together with the events of the fiducial (red points) and the full data set (orange points). We do not show error bars for the data, since observational uncertainties are always much smaller than the scatter induced by LSS and the host model. Note that the figure shows the DM after $\mathcal{D}_\mathrm{MW}$ and $\mathrm{D}_\mathrm{halo}$ have been subtracted. The likelihood at each redshift is normalised to peak at unity (to highlight the dynamical range). The dashed line shows the mean of the excess DM, i.e. the excess DM averaged using the likelihood. Therefore, the dashed line corresponds to the sum of the Macquart-relation, \Cref{eq:DM_LSS_avg}, and the host contribution at each redshift, $\langle\mathcal{D}_\mathrm{host}\rangle/(1+z)$.

Upon further inspection, we see that the dashed line moves closer to the peak of the likelihood with increasing redshift, indicating that the distribution becomes more symmetric, more Gaussian and higher redshifts. This is in line with what has been found in simulations \citep[e.g.][]{2024A&A...683A..71W,2025arXiv250707090K} and is as well-supported by several theoretical arguments: Let us assume for a moment that the LSS is a Gaussian random field. Then, the only non-Gaussian contribution comes from the log-normality (or rather non-Gaussianity) of the host distribution \citep{mcquinn_locating_2014,2024arXiv240308611T,2024ApJ...967...32M,2025OJAp....8E.127R}. As discussed earlier, the host contribution dominates the likelihood at low redshifts, leading to a large degree of non-Gaussianity and a mismatch between the mean and the mode of the likelihood. However, the LSS contribution itself is also not Gaussian due to structure formation. The expansion parameter controlling the perturbative treatment of structure growth is the relative fluctuation strength, $\Delta_\mathcal{D} = \sigma_\mathrm{LSS}/\langle\mathcal{D}_\mathrm{LSS}\rangle $. 

\begin{figure}
    \centering
    \includegraphics[width=0.45\textwidth]{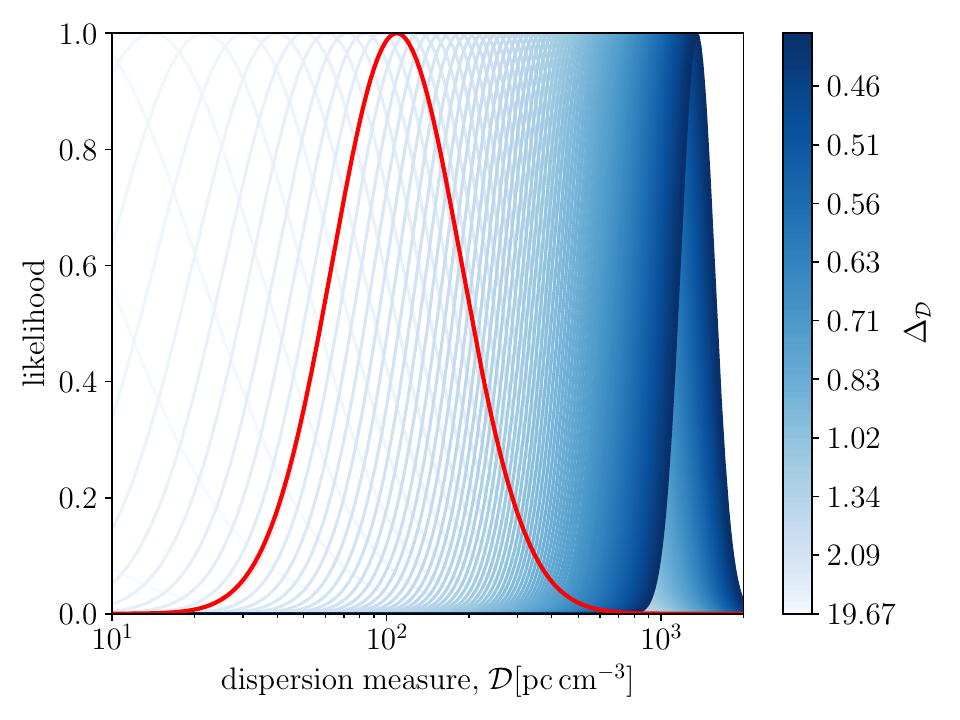}
    \caption{Evolution of the LSS likelihood (blue shades) in comparison to the host contribution at redshift zero (red curve). The colour bar indicates the fluctuation strength of the LSS relative to its mean $\Delta_\mathcal{D}$. Linear perturbation theory requires this to be well below unity. In that limit, one would expect a Gaussian likelihood.}
    \label{fig:fluctuation_strength}
\end{figure}

In \Cref{fig:fluctuation_strength}, we show the PDF of the DM as a function of $\Delta_\mathcal{D}$ (which traces redshift, $\Delta_\mathcal{D} = 0.5$ corresponds to a redshift of $0.2$ for our best fitting model). It is clear that the host contribution dominates at low redshifts, as discussed before. Furthermore, at higher redshifts, the PDF tends to become more Gaussian as $\Delta_\mathcal{D} \to 0$. In this limit, one can expand the log-normal distribution (or any distribution) around its peak in an asymptotic Edgeworth expansion and recover a Gaussian to leading order. The observed trend in \Cref{fig:likelihood} can therefore be well understood. It is noteworthy that $\Delta_\mathcal{D}$ drops below unity already at fairly low redshifts.

\begin{figure*}
    \centering
    \includegraphics[width=0.47\textwidth]{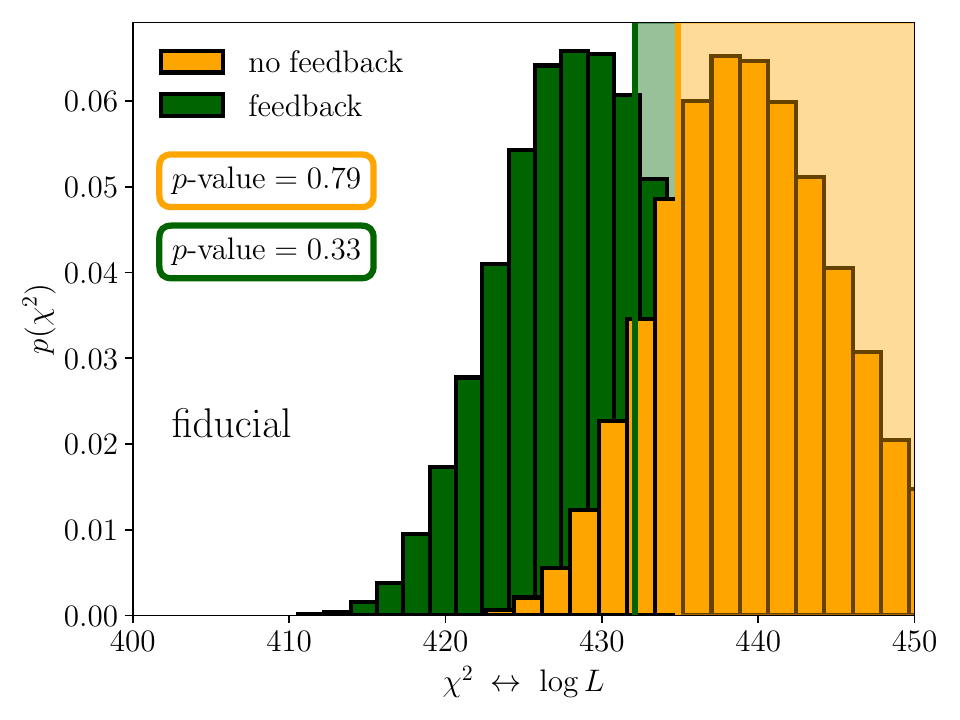}
    \includegraphics[width=0.47\textwidth]{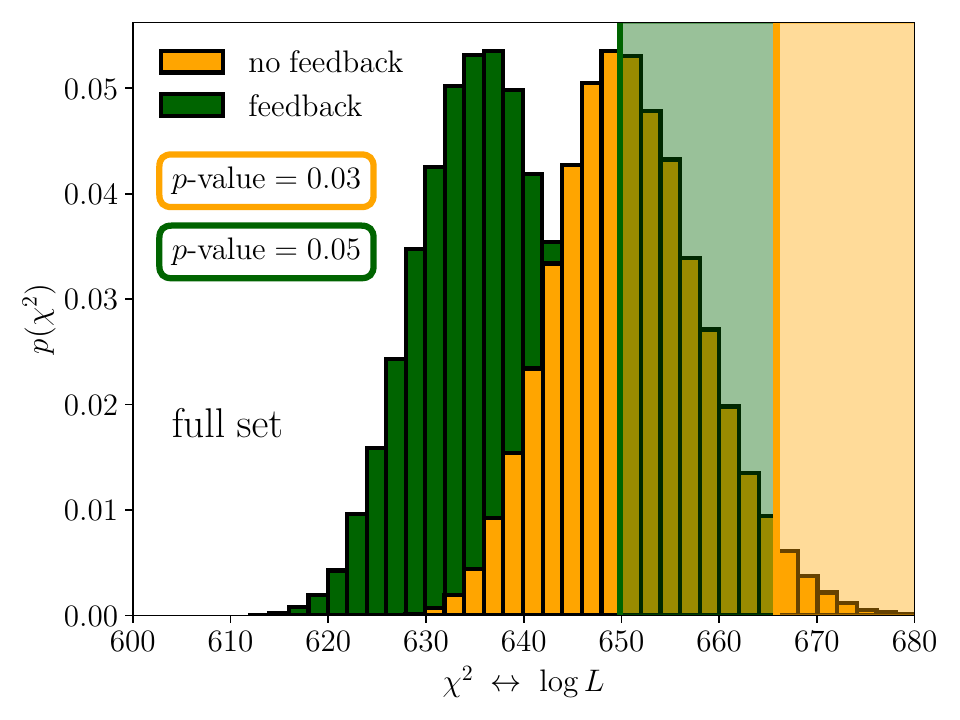}
    \caption{Probability distribution function of the $\chi^2$, which we identify as the log-likelihood. 
    The one-sided $p$-value, calculated via the probability-to-exceed, for the best fitting feedback (green) and the no feedback setting (orange). \textit{Left}: Fiducial sample of FRBs with both models being, in principle, a good fit to the data. \textit{Right}: Full sample, the fit is clearly poorer, and the feedback case fails the standard $p> 0.05$ test.}
    \label{fig:p_values}
\end{figure*}

\subsection{Goodness of fit}
\label{subsec:gof}
Next, we address the question of whether the models used are actually a good description of the data. This is being addressed in a Frequentist fashion. The standard approach is a $\chi^2$-test. Since the likelihood is, however, non-Gaussian, we cannot interpret it as a quadratic form in the same way as the $\chi^2$. Alternatively, one can draw $N$ random realisations of the data from the likelihood at the maximum of the posterior, $\boldsymbol{\theta}_0$, and calculate the best-fit log-likelihood, $\log L$, for each realisation. This serves as a proxy for the realisations $\chi^2_i$ and provides an estimate of its distribution based on many realisations of the data \citep{Tanseri:2022zfe}. By comparing the log-likelihood of the actual data, $\chi_\mathrm{data}$ to these realisations, we can estimate the one-sided $p$-value, or the probability-to-exceed as:
\begin{equation}
    p = P\left(\chi^2\geq\chi^{\mathstrut}_\mathrm{data}\right) \approx \sum_{\chi_i\leq\chi_\mathrm{data}}\frac{1}{N}\;.
\end{equation}
We show the resulting distribution and the corresponding $p$-values in \Cref{fig:p_values}. Here, the green histogram corresponds to our standard analysis including $T_\mathrm{AGN}$, while the orange histogram assumes no feedback. The coloured vertical lines indicate the actual log-likelihood of the data, and the shaded region indicates the area which contains the probability-to-exceed, which is also quoted in the colour-coded boxes. 
The left plot shows the fiducial data set, and the right plot corresponds to the full data set (the green and grey contours in \Cref{fig:corner} respectively). For the fiducial case, both cases with or without feedback are good fits, since the data is a very probable realisation of the respective models. The purely Frequentist point of view would therefore be that it is unnecessary to include the more complex model that includes baryonic feedback. From a Bayesian perspective, however, the situation is very different, since we have prior information that feedback exists in the Universe and that our models must allow for some modification \citep[see also][for a discussion in a very general setting]{gelman1996posterior}. This is, of course, reflected in the contours of \Cref{fig:corner}, and we will map those to the power spectrum suppression in the next section. It is, however, important to keep in mind that the current FRB data do not tell us whether feedback exists, as we do not know a priori that it does. As for every Bayesian analysis (especially in the case of limited constraining power), the results are subject to prior choices.

For the full set, i.e., the right plot of \Cref{fig:p_values}, the situation is somewhat different. Adding low-redshift FRBs drastically reduces the $p$-value in both cases, and the model without feedback even falls below the commonly used (but somewhat arbitrary) boundary of $p=0.05$. While this is not statistically significant in a strict sense, it suggests that modelling FRBs at low redshifts remains challenging. This will become even more important as more data, especially at low redshifts, become available. It may suggest that the Milky Way model, both for the ISM and the halo contributions, is too simplistic, or that the host model is not sufficiently flexible. Hence, future analyses may need to revise the assumptions to integrate FRB analysis into a cosmological setting \citep[see e.g.][for a baryonification model applied to the host contribution]{2025OJAp....8E.127R}.

\begin{figure*}
    \centering
    \includegraphics[width=0.9\textwidth,trim=.5cm 8cm 9.2cm .2cm, clip]{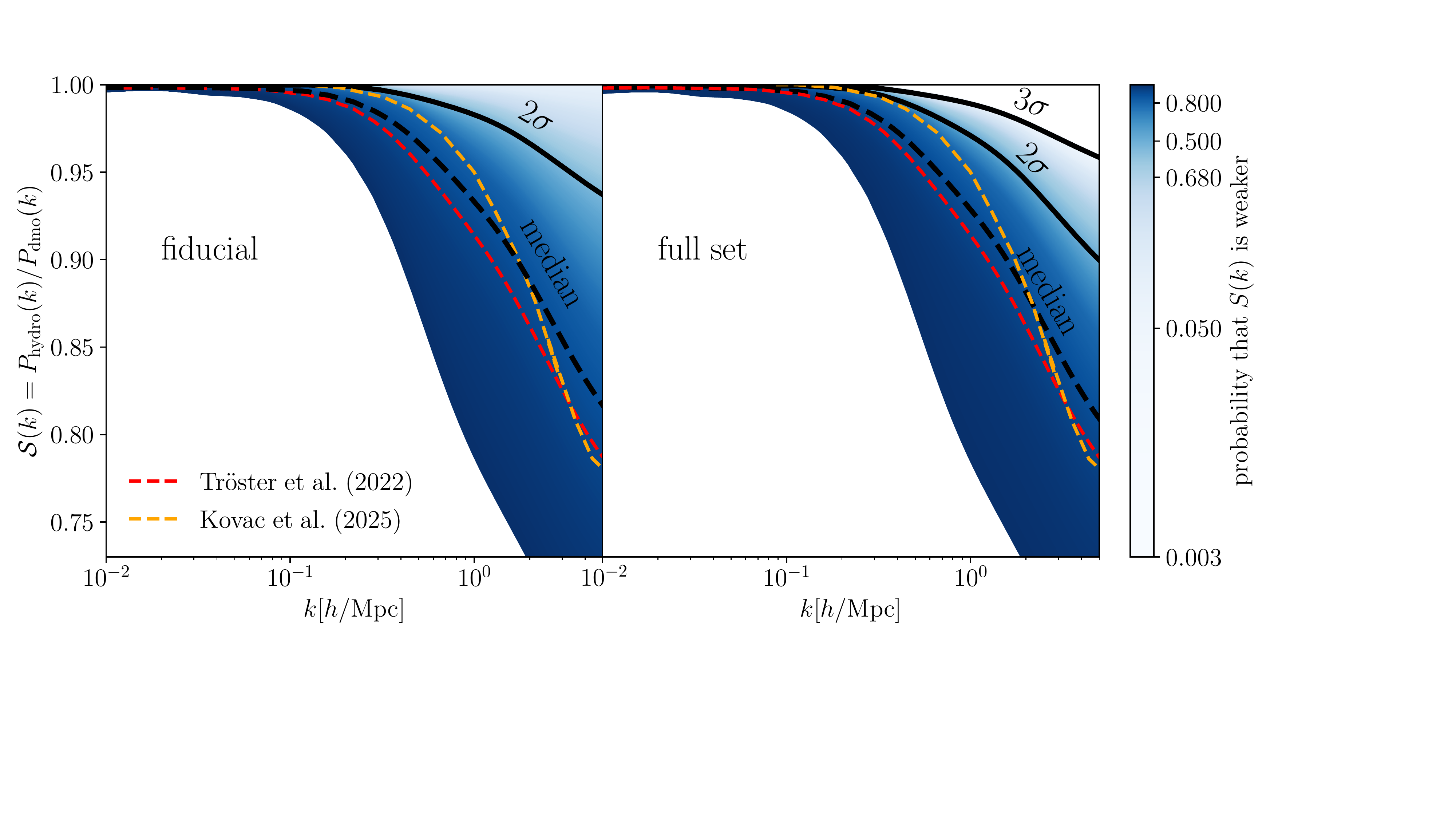}
    \caption{Suppression of the matter power spectrum as derived from the constraints shown in \Cref{fig:corner}. The black dashed line shows the median of the one-dimensional posterior distribution for $\log T_\mathrm{AGN}$. The $2\sigma$ line is shown as a solid black curve. As a colour-scale, we indicate the probability that feedback is weaker. This, of course, assumes that the \texttt{BAHAMAS} simulations provide the real feedback observed in the Universe. Additionally, the results from two other works are shown. In dashed red, we show the best fit measured by \citet{2022A&A...660A..27T} for the same model. In that study, the posterior is, however, very prior-dominated. The orange dashed line indicates the best fit obtained for the baryonification model \citep{2025arXiv250707892S} for $X$-Ray and kSZ data \citep{2025arXiv250707991K}.}
    \label{fig:suppression}
\end{figure*}

\subsection{Matter power spectrum suppression}
\label{subsec:sup}
As the last part of our analysis, we will translate the constraints on $\log T_\mathrm{AGN}$ from \Cref{fig:corner} into the resulting suppression of the matter power spectrum, which we define as
\begin{equation}
\mathcal{S}(k,z) = \frac{P_\mathrm{hydro}(k,z)}{P_\mathrm{dmo}(k,z)}\;,\quad \mathcal{S}(k) \equiv \mathcal{S}(k, 0)\;,
\end{equation}
where $P_\mathrm{hydro}(k,z)$ is the matter power spectrum from the hydrodynamical simulation, i.e. including baryons, and $P_\mathrm{dmo}(k,z)$ is the matter power spectrum in the same simulation but only with gravity, i.e. dark matter only (dmo). First, let us reiterate that AGN heating temperature ($T_\mathrm{AGN}$) correlates with feedback-driven modifications to halo profile concentration parameters for gas-poor and gas-rich haloes, halo-bound gas fractions, and the equation of state of the gas \citep{2020A&A...641A.130M}. Formally, it is related to the \texttt{BAHAMAS} subgrid heating parameter $\Delta T_\mathrm{heat}$, which governs the occurrence of AGN feedback such that it activates only after accumulating sufficient energy to heat a fixed number of gas particles by $\Delta T_\mathrm{heat}$. So while there is effectively no {dmo} setting, \citet{des/kids:2023} found that $\log T_\mathrm{AGN} = 7.3$ reproduces a dmo power spectrum, which we will use as the reference here. 

The main result of this paper can be seen in \Cref{fig:suppression}. We show $\mathcal{S}(k,0)$ for the fiducial and the full data set on the left and the right, respectively.
Here, the black dashed line marks the median of the one-dimensional marginal posterior of $T_\mathrm{AGN}$ (i.e. the sub-panel in \Cref{fig:corner}). Since we saw earlier that the constraints on $\log T_\mathrm{AGN}$ are slightly prior-dominated at the upper end, we will quote lower bounds. In other words, we specify the minimum feedback required by FRBs at a given confidence level.
The solid line shows the $2\sigma$ or $3\sigma$ lower bound on feedback. Put in another way, any $S(k,0)$ above that line and closer to unity is ruled out by $2\sigma$ or $3 \sigma$ respectively. Including the full data set, we find that the median and upper limit on feedback remain approximately the same; however, the lower limit on the feedback strength changes, and the dmo scenario is excluded by more than $3\sigma$. 

To put the constraints into numbers, we calculate the 68\% confidence interval of the marginal posterior distribution and find:
\begin{equation}
\begin{split}
    \log T_\mathrm{AGN} = &\;  {7.87^{\mathstrut+\hspace{.036cm}0.23}_{\mathstrut-0.27}} \quad \mathrm{fiducial}\\
  \log T_\mathrm{AGN} =&\;  {7.87^{\mathstrut+\hspace{.036cm}0.16}_{\mathstrut-0.22}} \quad \mathrm{full~set}\;,
\end{split}
\end{equation}
very consistent with the \texttt{BAHAMAS} model. For completeness, we also show the best fit value found in \citet{2022A&A...660A..27T} (dashed red curve), who assume the same model used here and fit the kSZ and cosmic shear cross-correlations. Their result is very consistent with ours, but they find weaker constraints in general and are partially prior-dominated, as the marginal posterior at the edge of the prior range dropped only to a tenth of the posterior value at the marginal mode. 

Additionally, we show in dashed orange recent results by \citet{2025arXiv250707991K} using $X$-Ray and kSZ data, but with a more flexible model introduced in \citet{2025arXiv250707892S}. Their results also exhibit fairly strong feedback, consistent with our findings and with the $f_{\mathrm{gas}}$–8$\sigma$ \texttt{FLAMINGO} simulation \citep{Schaye2023_flamingo,Schaller_pkFlamingo_2024}.

Lastly, \citet{2025arXiv250608932W} reported a detection of a cross-correlation between the DM and foreground galaxies at roughly $5\sigma$. While this is technically also a measurement of feedback, their results are difficult to interpret and cannot be uniquely attributed to feedback, as they are a combination of factors such as CHIME's beam size or non-linear galaxy bias.

\subsection{Possible limitations and discussion}
\label{subsec:limitations}
There are a couple of caveats to the analysis presented here, which we discuss. 

\begin{figure}
    \centering
    \includegraphics[width=0.45\textwidth]{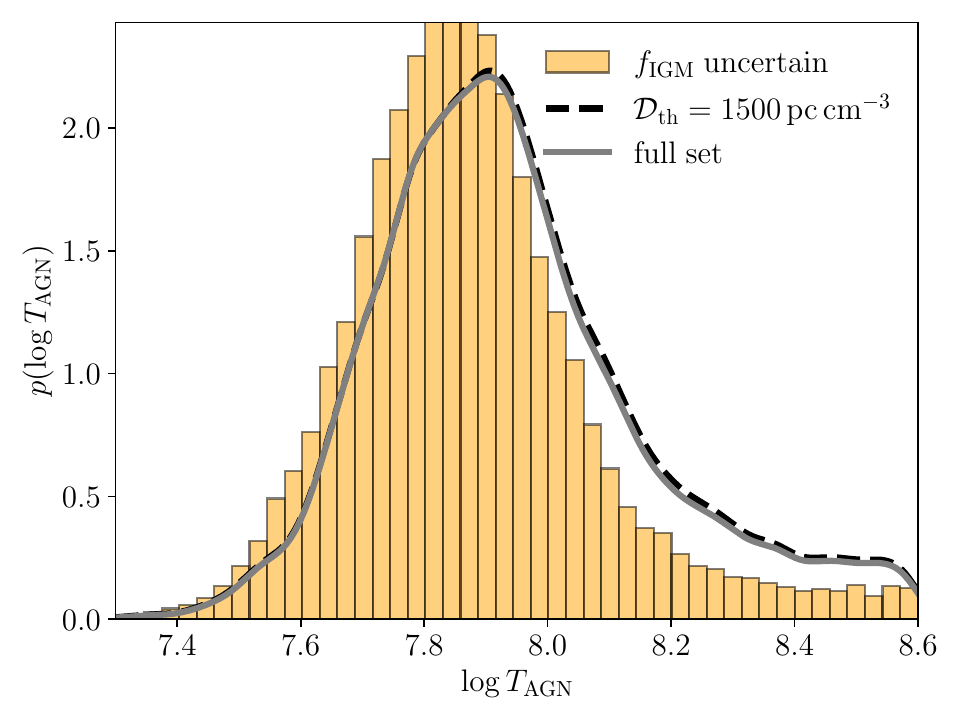}
    \caption{\review{Marginalised posterior distribution of the feedback strength $\log T_\mathrm{AGN}$. The grey curve corresponds to the constraints from the full set (see also grey curves and contours in \Cref{fig:corner}). The black-dashed curve includes a selection effect with a hard cut-off in the observed dispersion measure. The orange histogram additionally marginalises over the uncertainty of $f_\mathrm{IGM}$.}}
    \label{fig:selction}
\end{figure}

\review{
\subsubsection{Selection}
First, there is the issue of completeness of the survey, or in other words, the luminosity function of FRBs \citep[e.g.][]{2022MNRAS.516.4862J,2025PASA...42...17H,2025ApJ...986..100G}. This may introduce selection effects in the FRB sample, particularly for large DM values. Since this selection will preferentially select FRBs with low DMs, as they are more likely to be detected, provided an overall sensitivity of a given telescope, the corresponding sample of FRBs will prefer FRBs which are scattered low in the DM-$z$ relation, thus biasing the inferred scatter low. \citet{2025arXiv251116850S} demonstrated that selection effects affect current sample sizes only to a negligible degree. For the analysis done here, we carry out a similar test, setting a threshold for the maximum observed DM $\mathcal{D}_\mathrm{th}$. We then rerun our analysis on the full set, ensuring that $p(\mathcal{D}_\mathrm{tot}> \mathcal{D}_\mathrm{th}) = 0$, with $\mathcal{D}_\mathrm{th} = 1500\,\mathrm{pc}\,\mathrm{cm}^{-3} $. \Cref{fig:selction} demonstrates that this effect is very small. The grey line shows the $\mathcal{D}_\mathrm{th} \to \infty$, for reference. }

\review{
 \subsubsection{$f_\mathrm{IGM}$ uncertainty}
Our fiducial analysis assumes that $f_\mathrm{IGM}$ is known exactly given the prescription in \citet{2020Natur.581..391M}. To allow for uncertainties in $f_\mathrm{IGM}$, we rescale its value by a free nuisance parameter $\mathcal{F}$ such that
\begin{equation}
f_{\mathrm{IGM,eff}}(z) = \min \left[ 1,\; f_\mathrm{IGM}(z)\mathcal{F} \right].
\end{equation}
By marginalising over $\mathcal{F}$, we propagate the residual uncertainty in $f_\mathrm{IGM}$ into $\log T_\mathrm{AGN}$. We do not assume any additional prior apart from the one introduced above. \Cref{fig:selction} shows the resulting posterior as the orange histogram. The probability mass shifts towards lower $\log T_\mathrm{AGN}$, indicating slightly weaker feedback:
\begin{equation*}
\log T_\mathrm{AGN} = {7.87^{\mathstrut+\hspace{.036cm}0.17}_{\mathstrut-0.16}}\quad\mathrm{full set,\;}\mathcal{F}\;\mathrm{marginalised.}
\end{equation*}
Overall, the limits on the feedback strength are largely unaffected. The reason for this is that $\mathcal{F}$ is constrained by the DM-redshift relation since the other cosmological parameters are effectively fixed by the Planck prior. Hence, the DM-redshift largely fits $\mathcal{F}$ and the two host contribution parameters, freeing up sensitivity to $\log T_\mathrm{AGN}$ when the scatter in the DM-redshift relation.}

\begin{figure}
    \centering
    \includegraphics[width=0.45\textwidth]{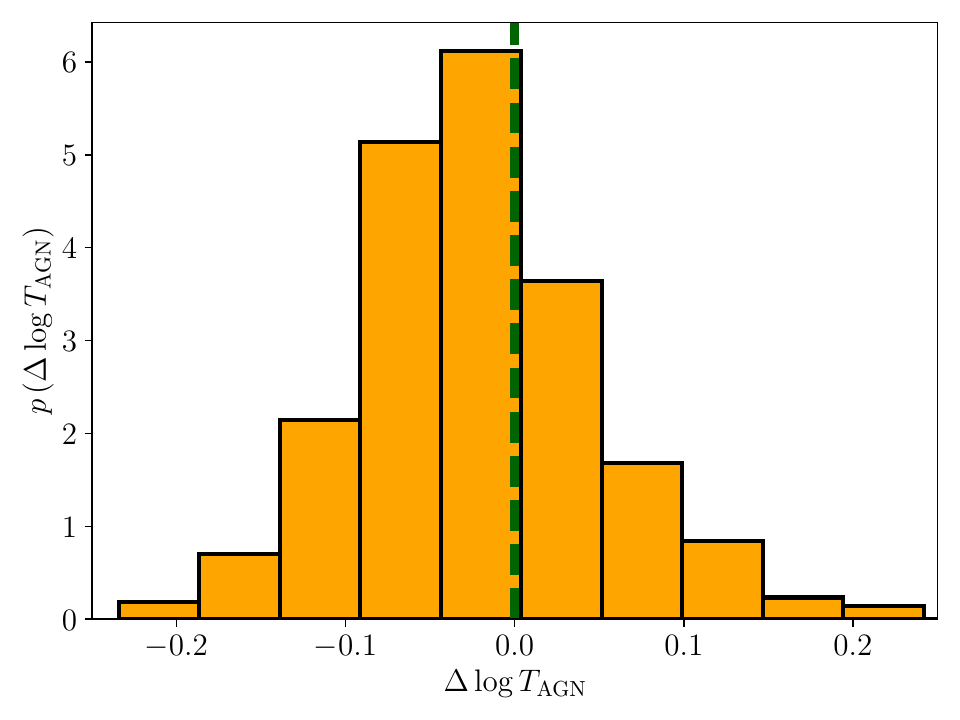}
    \caption{Histogram of the maximum posterior value of $\log T_\mathrm{AGN}$ of the same number of FRBs as in the fiducial set, but randomly sampled from the full set. The histogram has been shifted so that the zero line corresponds to the scenario from the fiducial case.}
    \label{fig:shuffl}
\end{figure}

Secondly, as discussed earlier, the fact that low-redshift FRBs will dominate the sample requires a very accurate modelling of both the Milky Way and the host contribution. The presented results indicate, however, that the first point is not a dominating systematic at the moment, as splitting the sample does not yield any systematic shift in the best-fitting values but rather changes the measurement uncertainties. We show an attempt of quantifying the overall scatter in $\log T_\mathrm{AGN}$ in \Cref{fig:shuffl}. To obtain this plot, we use a random sample of the full set of FRBs, with the same size as the fiducial set. Next, the posterior is maximised, and we report the $\log T_\mathrm{AGN}$ value, upon repeating this exercise many times, we arrive at \Cref{fig:shuffl} where $\log T_\mathrm{AGN}$ was shifted such that zero falls onto the best fit fiducial case,  $\log T_\mathrm{AGN} = 7.87$. It is clear that the fiducial data set is not a very unlikely draw and that selection effects would have biased our result. This is, of course, only true for the available FRBs and merely shows that the fiducial case was not particularly special relative to the full data set.

At the same time, the $p$-value issue discussed in Section \ref{subsec:gof} suggests that the modelling of the host and the Milky Way will require some refinement in the future. 

Another limitation is the model for the feedback itself. While the power spectrum suppression is representative of the \texttt{BAHAMAS} via variations of both cosmological parameters and $\log T_\mathrm{AGN}$, this one-parameter family has a limited functional form. Put differently, while other emulators, e.g. \citet{Giri2021_BCemu}, can reproduce the \texttt{BAHAMAS} suppression, the model here does not cover all kinds of suppressions as measured in hydrodynamical simulations \citep[see e.g.][for a list of different feedback implementations]{2021ApJ...915...71V}. Nonetheless, since $\log T_\mathrm{AGN}$ is sufficient to describe current weak lensing data \citep{wright_kids_2025} and the constraints presented here show that FRBs indeed require relatively strong feedback.

\review{
Very recently, \citet{2025arXiv250716816L} have investigated the relationship between the host DM and the host's stellar mass, $M_\star$, for a sample of 20 FRBs. They find a negative trend of $\mathrm{D}_\mathrm{host}$ with $M_\star$, as observed in the low-mass halos (with stellar masses well below $10^{11}\, M_\odot$). 
This finding is at first somewhat counterintuitive if one assumes a simple scaling of gas content with stellar mass. Therefore, it is suggested that efficient baryonic feedback mechanisms expel gas from these more massive galaxies.
This contrasts with the relationship observed in some CAMELS sub-grid models, in particular Astrid \citep{2021ApJ...915...71V,2024arXiv240308611T,2024ApJ...967...32M}. By translating this relationship into a suppression of the power spectrum, we find that small suppressions, $\mathrm{max}(\mathcal{S}(k,0)) > 0.95$, are excluded, consistent with what is presented here. We emphasise that the two papers rely on different observables. We model the effect of feedback on the distribution of electrons within the LSS and therefore on the scatter of $\mathcal{D}_\mathrm{LSS}$, while \citet{2025arXiv250716816L} focus on $\mathcal{D}_\mathrm{host}$. While their analysis is not plagued by selection effects as it relies entirely on local bursts with low DM, a small caveat is the assumption of a Gaussian likelihood for $\mathcal{D}_\mathrm{host}$. Even with the conservative selection employed there, one would expect $\mathcal{D}_\mathrm{host}$ to follow a long-tailed distribution.}

\section{Conclusions}
\label{sec:conclusions}
\label{sec:conclusions}
In this work, we constrain baryonic feedback for the first time using the scatter in the dispersion measure - redshift relation of FRBs. We forward-model the power spectrum of the hot (ionised) gas using a halo model approach, which has been calibrated to the \texttt{BAHAMAS} simulations, providing a one-parameter model for the strength of baryonic feedback which enables us to sweep a range of possible feedback strengths consistent with the \texttt{BAHAMAS} suite. 

By assuming a prior on the cosmological parameters given by Planck \citep{planck_collaboration_planck_2020}, we can marginalise over the host contribution of FRBs and find constraints on baryonic feedback and map it to the suppression of the matter power spectrum. Using a set of $69$ FRBs (fiducial data set) and an expanded (full) set with $28$ additional FRBs, with most of them at $z<0.1$, we constrain feedback and obtain the following main conclusions:
\begin{enumerate}
    \item A scenario with no baryonic feedback, i.e. where the power spectrum suppression is unity, is ruled out at 99.7\% confidence for the fiducial data set. In the case of the extended data set, we can exclude a dark-matter-only scenario with $5\sigma$. 

\item The median of the suppression measured by our model is in line with strong-feedback predictions from the \texttt{FLAMINGO} simulations. 
However, with the current data, the results are still consistent with the fiducial run of \texttt{FLAMINGO}. Nonetheless, the results indicate stronger feedback and align well with other works \citep[e.g.][]{Bigwood2024, McCarthy_kSZ2024,2025arXiv250707991K}.

\item We test the importance of the Milky Way contribution by marginalising over it and find that the influence on the constraints of feedback are negligible.

\item With an increasing number of FRBs, correlations between different sightlines will become more and more important \citep{reischke_cosmological_2023}. This issue is addressed by calculating the Gaussian covariance between the bursts and removing the highly correlated ones with a correlation coefficient $r\geq 0.3$. In total, 4 such pairs are identified and removed. The current amount of correlation does not warrant further investigation. With the next jump in sensitivity, this effect needs to be taken into account \citep[see][for details]{reischke_cosmological_2023,2024arXiv241007084K} to deliver unbiased results. 

\end{enumerate}
We thus conclude that the constraints presented here are robust under the current modelling assumptions, and that FRBs indeed prefer strong feedback, as do other probes of the baryon distribution. The measurement, therefore, provides the first step toward using FRBs as a powerful cosmological probe in the era of precision cosmology with LSS.

\section*{Acknowledgements}
We thank an anonymous referee for helping make the presented analysis more robust.
RR would like to thank Stella Ocker for the correspondence regarding the MW contribution. We thank Kritti Sharma for insightful discussions.
SH was supported by the Excellence Cluster ORIGINS, which is funded by the Deutsche Forschungsgemeinschaft (DFG, German Research Foundation) under Germany’s Excellence Strategy - EXC-2094 - 390783311. We used \texttt{matplotlib} \citep{Hunter_matplotlib_2007} for the standard plots and \texttt{getdist} \citep{2019arXiv191013970L} for contour plots. A lot of the computations were done with the help of \texttt{SciPy} \citep{2020SciPy-NMeth} and \texttt{NumPy} \citep{harris2020array}. 

\bibliographystyle{mnras}
\bibliography{more_bib} 

\renewcommand\thesection{\alpha{section}}
\renewcommand\thesubsection{\thesection.\arabic{subsection}}
\renewcommand\thefigure{\thesection.\arabic{figure}} 
\setcounter{figure}{0}
\newpage
\newpage

\renewcommand\thesection{\alpha{section}}
\renewcommand\thesubsection{\thesection.\arabic{subsection}}
\renewcommand\thefigure{\thesection.\arabic{figure}} 
\setcounter{figure}{0}

\cleardoublepage
\appendix

\section{Influence of the likelihood assumption}\label{app:likelihood}
\begin{figure}
    \centering
    \includegraphics[width=0.5\linewidth]{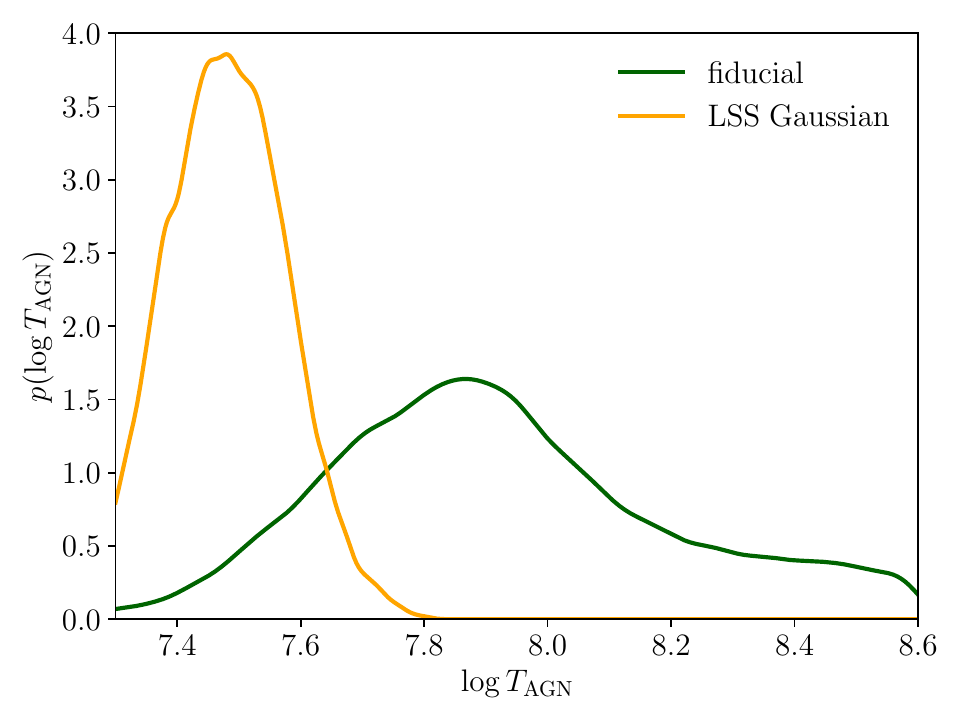}
    \caption{In green, we show the fiducial analysis (also green curve in \Cref{fig:corner}). The yellow curve is the result of replacing $p_\mathrm{LSS}$ by a Gaussian with the same variance in \Cref{eq:likelihood:full}. All other parameters have been marginalised over.}
    \label{fig:gaussian_likelihood}
\end{figure}

We show the influence of the assumption of a Gaussian likelihood on the inferred feedback strength. In particular, we replace $p_\mathrm{LSS}$ by a Gaussian with the same variance in \Cref{eq:likelihood:full} and redo the inference. In \Cref{fig:gaussian_likelihood}, one can see that the assumption of a Gaussian likelihood introduces an underestimation of the feedback strength. The reason for this is that the model attempts to fit the long log-normal tail of the $\mathcal{D}_\mathrm{LSS}$ contribution by increasing the variance of the LSS, $\mathrm{C}_{ii}$. This can only be done by lowering $\log T_\mathrm{AGN}$ and hence the strength of feedback. 

\section{Data}\label{app:data}
We make use of publicly available FRBs with host identification and measured redshifts. We summarise the data used in this work in \Cref{tab:app_frb_list} for the fiducial data set and the additional events at low redshifts making up the full data set in \Cref{tab:app_frb_list_2}. 
\begin{table}
    \centering
    \begingroup
    \renewcommand{\arraystretch}{1}
    \tiny
    \begin{tabular}{lcccccc}
    \hline
\hline
FRB\hspace{1cm} & $\hspace{.5cm}\mathcal{D}_\mathrm{obs}[\mathrm{pc\;cm}^{-3}]$\hspace{.5cm} & \hspace{.5cm}$z$\hspace{.5cm} & \hspace{.5cm}$\mathcal{D}_\mathrm{MW}[\mathrm{pc\;cm}^{-3}]$\hspace{.5cm} & \hspace{.5cm}ra[rad]\hspace{.5cm} & \hspace{.5cm}dec[rad]\hspace{.5cm} & \hspace{.5cm}Source \hspace{.5cm}\\
20230708A  &  411.51  &  0.105  &  50.0  &  -2.7889  &  -0.9661 & \citet{Shannon_202307} \\ 
20191106C  &  332.2  &  0.10775  &  25.0  &  -2.9094  &  0.7505 & \citet{chime_2021_first}\\ 
20220914A  &  631.28  &  0.1139  &  55.2  &  -2.8134  &  1.28 & \citet{sherman_2024_deep} \\ 
20190608  &  339.5  &  0.11778  &  37.2  &  -2.7529  &  -0.1378 & \citet{zwaniga_chimefrb_discovery_2021} \\ 
20190110C  &  221.6  &  0.12244  &  37.1  &  -2.8515  &  0.7233 & \citet{zwaniga_chimefrb_discovery_2021} \\ 
20240310A  &  601.8  &  0.127  &  36.0  &  -3.1182  &  -0.7756 & \citet{Shannon_20240712} \\ 
20240213A  &  357.4  &  0.1185  &  32.1  &  -2.9482  &  1.2929 & \citet{law_dsa_20240601} \\ 
20230628A  &  345.15  &  0.1265  &  30.83  &  -2.9475  &  1.2616 & \citet{law_dsa_20240601} \\ 
20210807D  &  251.3  &  0.1293  &  121.2  &  -2.7934  &  -0.0133 & \citet{2022MNRAS.516.4862J}\\ 
20240114A  &  527.7  &  0.13  &  38.82  &  -2.767  &  0.0756 & \citet{abbott_chimefrb_20240202} \\ 
20210410D  &  578.78  &  0.1415  &  56.2  &  -2.7622  &  -1.3844 & \citet{caleb_2023_subarcsec} \\ 
20231226A  &  329.9  &  0.1569  &  145.0  &  -2.9608  &  0.1066 & \citet{Shannon_craft_20240112} \\ 
20230526A  &  361.4  &  0.157  &  50.0  &  -3.1157  &  -0.9201 & \citet{shannon_20230527} \\ 
20220920A  &  314.99  &  0.158239  &  40.3  &  -2.862  &  1.2378 & \citet{sherman_2024_deep} \\ 
20200430  &  380.25  &  0.16  &  27.0  &  -2.8743  &  0.216 & \citet{kumar_2021_craft} \\ 
20210603A  &  500.147  &  0.177  &  40.0  &  -3.1296  &  0.3705 & \citet{leung_2022_oneoff} \\ 
20220529A  &  246.0  &  0.1839  &  30.92  &  -3.1194  &  0.3601 & \citet{li_activerepeating_2025} \\ 
20220725A  &  290.4  &  0.1926  &  31.0  &  -2.7305  &  -0.6281 & \citet{shannon_craft_20220726} \\ 
20121102  &  557.0  &  0.1927  &  188.0  &  -3.045  &  0.5783 & \citet{petroff_frbcat_20200812} \\ 
20221106A  &  343.8  &  0.2044  &  35.0  &  -3.0756  &  -0.4463 & \citet{shannon_craft_20221113} \\ 
20240215A  &  549.5  &  0.21  &  42.8  &  -2.8292  &  1.2258  & \citet{law_dsa_20240601} \\ 
20210117A  &  728.95  &  0.214  &  34.4  &  -2.746  &  -0.2819 & \citet{shannon_craft_20210119} \\ 
20221027A  &  452.5  &  0.229  &  40.59  &  -2.9893  &  1.2584 & \citet{law_dsa_20230810} \\ 
20191001  &  507.9  &  0.234  &  44.7  &  -2.7654  &  -0.9555 & \citet{shannon_2019_ASKAP191001} \\ 
20190714  &  504.13  &  0.2365  &  39.0  &  -2.9275  &  -0.2273 & \citet{shannon_2019_ASKAP190714} \\ 
20221101B  &  490.7  &  0.2395  &  192.35  &  -2.7434  &  1.2336 & \citet{law_dsa_20230810}\\ 
\textcolor{red}{20190520B}  &  1210.3  &  0.241  &  113.0  &  -2.8617  &  -0.197 & \citet{niu_transient_20210925} \\ 
20220825A  &  651.24  &  0.241397  &  79.7  &  -2.7786  &  1.2668 & \citet{sherman_2024_deep} \\ 
20191228  &  297.5  &  0.2432  &  33.0  &  -2.7408  &  -0.5165 & \citet{petroff_frbcat_20200812} \\ 
20220307B  &  499.27  &  0.248123  &  135.7  &  -2.7333  &  1.26 & \citet{sherman_2024_deep} \\ 
20221113A  &  411.4  &  0.2505  &  115.37  &  -3.0585  &  1.2271 & \citet{law_dsa_20240601} \\ 
\textcolor{red}{20220831A}  &  1146.25  &  0.262  &  187.94  &  -2.7475  &  1.2258 & \citet{law_dsa_20230810} \\ 
20231123B  &  396.7  &  0.2625  &  33.81  &  -2.8594  &  1.2354 & \citet{law_dsa_20240601} \\ 
20230307A  &  608.9  &  0.271  &  29.47  &  -2.9347  &  1.2513 & \citet{law_dsa_20240601} \\ 
20221116A  &  640.6  &  0.2764  &  196.17  &  -3.1169  &  1.2681 & \citet{law_dsa_20240601} \\ 
20220105A  &  580.0  &  0.2785  &  22.0  &  -2.8986  &  0.3921 & \citet{shannon_craft_20220126} \\ 
20210320C  &  384.8  &  0.2797  &  42.0  &  -2.9037  &  -0.2814 & \citet{shannon_craft_20230204} \\ 
20221012A  &  441.08  &  0.284669  &  54.4  &  -2.8149  &  1.2309 & \citet{sherman_2024_deep} \\ 
20240229A  &  491.15  &  0.287  &  29.52  &  -2.9438  &  1.2335 & \citet{law_dsa_20240402} \\ 
20190102  &  364.5  &  0.2913  &  57.3  &  -2.7664  &  -1.3871 & \citet{2020Natur.581..391M} \\ 
20220506D  &  396.97  &  0.30039  &  89.1  &  -2.7715  &  1.2711 & \citet{sherman_2024_deep} \\ 
20230501A  &  532.5  &  0.301  &  180.18  &  -2.746  &  1.2378 & \citet{law_dsa_20240601} \\ 
20180924  &  361.42  &  0.3214  &  40.5  &  -2.7622  &  -0.7138 & \citet{2019Sci...365..565B} \\ 
20230626A  &  451.2  &  0.327  &  32.51  &  -2.8674  &  1.2415 & \citet{law_dsa_20240601} \\ 
20180301  &  536.0  &  0.3304  &  152.0  &  -3.0331  &  0.0815 & \citet{petroff_frbcat_20200812} \\ 
20231220A  &  491.2  &  0.3355  &  44.55  &  -2.9974  &  1.2856 & \citet{law_dsa_20240601} \\ 
20211203C  &  635.0  &  0.3439  &  63.0  &  -2.9036  &  -0.5477 & \citet{shannon_craft_20230204} \\ 
20220208A  &  437.0  &  0.351  &  128.79  &  -2.7663  &  1.2224 & \citet{law_dsa_20230810} \\ 
20220726A  &  686.55  &  0.361  &  111.42  &  -3.0556  &  1.2205 & \citet{law_dsa_20230810} \\ 
20230902A  &  440.1  &  0.3619  &  34.0  &  -3.0809  &  -0.8261 & \citet{shannon_craft_20230902} \\ 
20200906  &  577.8  &  0.3688  &  36.0  &  -3.0881  &  -0.2458 & \citet{2022AJ....163...69B} \\ 
20240119A  &  483.1  &  0.37  &  30.98  &  -2.8804  &  1.2499 & \citet{law_dsa_20240601} \\ 
20220330D  &  468.1  &  0.3714  &  57.83  &  -2.9511  &  1.2279 & \citet{law_dsa_20230810} \\ 
20190611  &  321.4  &  0.378  &  43.67  &  -2.7684  &  -1.3857 & \citet{2020Natur.581..391M} \\ 
20220501C  &  449.5  &  0.381  &  31.0  &  -2.7316  &  -0.5671 & \citet{law_dsa_20230810} \\ 
20220204A  &  612.2  &  0.4  &  46.02  &  -2.8225  &  1.2169 & \citet{law_dsa_20230810} \\ 
20230712A  &  586.96  &  0.4525  &  30.93  &  -2.9469  &  1.2664 & \citet{law_dsa_20240601} \\ 
20181112  &  589.27  &  0.4755  &  102.0  &  -2.7607  &  -0.9245 & \citet{2019Sci...366..231P} \\ 
20220310F  &  462.24  &  0.477958  &  45.4  &  -2.9848  &  1.2827 & \citet{sherman_2024_deep} \\ 
20220918A  &  656.8  &  0.491  &  41.0  &  -3.1211  &  -1.2359 & \citet{shannon_craft_20220919} \\ 
20190711  &  593.1  &  0.522  &  56.4  &  -2.1349  &  -1.4025 & \citet{petroff_frbcat_20200812} \\ 
20230216A  &  828.0  &  0.531  &  27.05  &  -2.9595  &  0.06 & \citet{law_dsa_20240601} \\ 
20230814A  &  696.4  &  0.5535  &  137.83  &  -2.7507  &  1.2745 & \citet{ravi_dsa_20230816}\\ 
20221219A  &  706.7  &  0.554  &  38.59  &  -2.8418  &  1.2501 & \citet{law_dsa_20240601} \\ 
20190614  &  959.2  &  0.6  &  83.5  &  -3.0659  &  1.2864 & \citet{zwaniga_chimefrb_discovery_2021} \\ 
20220418A  &  623.25  &  0.622  &  37.6  &  -2.8867  &  1.2234 & \citet{sherman_2024_deep} \\ 
20190523  &  760.8  &  0.66  &  37.0  &  -2.9007  &  1.2648 & \citet{2019Natur.572..352R} \\ 
20240123A  &  1462.0  &  0.968  &  113.01  &  -3.0622  &  1.2557 & \citet{law_dsa_20240601} \\ 
20221029A  &  1391.05  &  0.975  &  36.4  &  -2.9764  &  1.2645 & \citet{law_dsa_20230810} \\ 
20220610A  &  1457.624  &  1.016  &  31.0  &  -2.7331  &  -0.5849 & \citet{shannon_craft_20220612} \\ 
20230521B  &  1342.9  &  1.354  &  209.66  &  -2.7331  &  1.2416 & \citet{law_dsa_20240601} \\  
\end{tabular}
\endgroup
    \caption{The collection of 71 FRBs in the \textbf{fiducial} data set. The two FRBs marked in red show a very high intrinsic DM and are therefore excluded from the analysis as explained in the main text.}
\label{tab:app_frb_list}
\end{table}

\begin{table}[]
    \centering
    \begingroup
    \renewcommand{\arraystretch}{1.}
    \tiny
    \begin{tabular}{lcccccc}
    \hline
\hline
FRB\hspace{1cm} & $\hspace{.5cm}\mathcal{D}_\mathrm{obs}[\mathrm{pc\;cm}^{-3}]$\hspace{.5cm} & \hspace{.5cm}$z$\hspace{.5cm} & \hspace{.5cm}$\mathcal{D}_\mathrm{MW}[\mathrm{pc\;cm}^{-3}]$\hspace{.5cm} & \hspace{.5cm}ra[rad]\hspace{.5cm} & \hspace{.5cm}dec[rad]\hspace{.5cm} & \hspace{.5cm}Source \hspace{.5cm}\\

FRB 20181223C & 111.61 & 0.03024 & 19.9 & 3.1576 & 0.4801 & \citet{2024ApJ...971L..51B} \\
FRB 20190303A & 223.2 & 0.064 & 29.8 & 3.6237 & 0.8400 & \citet{2023ApJ...950..134M} \\
FRB 20190418A & 182.78 & 0.07132 & 70.2 & 1.1484 & 0.2808 & \citet{2024ApJ...971L..51B} \\
FRB 20200223B & 201.8 & 0.06024 & 45.6 & 0.1440 & 0.5032 & \citet{2024ApJ...961...99I} \\
FRB 20201124A & 413.52 & 0.098 & 139.9 & 1.3464 & 0.4548 & \citet{2022MNRAS.513..982R} \\
FRB 20211127I & 234.83 & 0.0469 & 42.5 & 3.4873 & -0.3292 & \citet{2023ApJ...954...80G} \\
FRB 20211212A & 206 & 0.0715 & 27.1 & 2.7458 & 0.0292 & \citet{2023ApJ...954...80G} \\
FRB 20220717A & 637.34 & 0.36295 & 118.3 & 5.1097 & -0.3365 & \citet{2024MNRAS.532.3881R} \\
FRB 20230124A & 590.574 & 0.0939 & 38.6 & 4.0425 & 1.2426 & \citet{2024Natur.635...61S} \\
FRB 20230203A & 420.1 & 0.1464 & 67.299 & 1.0168 & 0.6224 & \citet{2025arXiv250211217C} \\
FRB 20230222A & 706.1 & 0.1223 & 33.31 & 1.8681 & 0.1959 & \citet{2025arXiv250211217C} \\
FRB 20230222B & 187.8 & 0.11 & 56.9977 & 4.1687 & 0.5392 & \citet{2025arXiv250211217C} \\
FRB 20230311A & 364.3 & 0.1918 & 67.2352 & 1.5916 & 0.9765 & \citet{2025arXiv250211217C} \\
FRB 20230703A & 291.3 & 0.1184 & 38.1566 & 3.2214 & 0.8511 & \citet{2025arXiv250211217C} \\
FRB 20230730A & 312.5 & 0.2115 & 114.6465 & 0.9549 & 0.5784 & \citet{2025arXiv250211217C} \\
FRB 20230808F & 653.2 & 0.3472 & 31 & 0.9302 & -0.9060 & \citet{2025MNRAS.538.1800H} \\
FRB 20230930A & 456 & 0.0925 & 70 & 0.1832 & 0.7212 & \citet{2025arXiv250302947A} \\
FRB 20231011A & 186.3 & 0.0783 & 58.5524 & 0.3189 & 0.7289 & \citet{2025arXiv250211217C} \\
FRB 20231017A & 344.2 & 0.245 & 31.1515 & 6.0502 & 0.6388 & \citet{2025arXiv250211217C} \\
FRB 20231025B & 368.7 & 0.3238 & 65.5787 & 4.7356 & 1.1175 & \citet{2025arXiv250211217C} \\
FRB 20231120A & 438.9 & 0.07 & 43.8 & 2.5072 & 1.2526 & \citet{2024Natur.635...61S} \\
FRB 20231123A & 302.1 & 0.0729 & 37.1085 & 1.4404 & 0.0779 & \citet{2025arXiv250211217C} \\
FRB 20231204A & 221 & 0.0644 & 34.9372 & 3.6237 & 0.8400 & \citet{2025arXiv250211217C} \\
FRB 20231206A & 457.7 & 0.0659 & 37.5114 & 1.9636 & 0.9819 & \citet{2025arXiv250211217C} \\
FRB 20231229A & 198.5 & 0.019 & 65.1335 & 0.4617 & 0.6127 & \citet{2025arXiv250211217C} \\
FRB 20231230A & 131.4 & 0.0298 & 32.2305 & 1.2707 & 0.0418 & \citet{2025arXiv250211217C} \\
FRB 20240201A & 374.5 & 0.042729 & 38.6 & 2.6168 & 0.2459 & \citet{2025PASA...42...36S} \\
FRB 20240209A & 176.518 & 0.1384 & 63.1 & 5.0568 & 1.5005 & \citet{2025ApJ...979L..22E} \\
\end{tabular}
\endgroup
    \caption{The collection of 28 additional FRBs. Together with the ones presented in \Cref{tab:app_frb_list}, these make up the \textbf{full set} of FRBs used in this study.}
\label{tab:app_frb_list_2}
\end{table}

\end{document}